\documentclass[%
 reprint,        
 amsmath,amssymb,
 aps,            
 prl
]{revtex4-2}

\usepackage{graphicx}   
\usepackage{dcolumn}    
\usepackage{bm}         
\usepackage{braket}     
\usepackage{hyperref}   
\usepackage{siunitx}    
\usepackage{epstopdf}
\usepackage{color}

\newcommand{\response}[1]{{\color{black}{#1}}}

\DeclareSIUnit\bar{bar}
\DeclareSIUnit{\molar}{M}

\hypersetup{
    colorlinks=true,
    linkcolor=blue,
    citecolor=blue,
    urlcolor=blue
}

\begin{document}

\title{Exciton-mediated optical control of liquid-solid friction}

\author{Timur Pryadilin\textsuperscript{1,2}}
\author{Alexey Kavokin\textsuperscript{1,2,3,4}}
\author{Baptiste Coquinot\textsuperscript{5}}
\email{baptiste.coquinot@ista.ac.at}

\affiliation{\textsuperscript{1}Abrikosov Center for Theoretical Physics, MIPT, Dolgoprudnyi, Moscow Region 141701, Russia}
\affiliation{\textsuperscript{2}Russian Quantum Center, Skolkovo, Moscow, 121205, Russia}
\affiliation{\textsuperscript{3}School of Science, Westlake University, 18 Shilongshan Road, Hangzhou 310024, Zhejiang Province, China}
\affiliation{\textsuperscript{4}Department of Physics, St. Petersburg State University, University Embankment, 7/9, St. Petersburg, 199034, Russia}
\affiliation{\textsuperscript{5}Institute of Science and Technology Austria (ISTA), Am Campus 1, 3400 Klosterneuburg, Austria}

\date{\today}

\begin{abstract}
Interfacial friction in nanofluidic systems can arise from fluctuation-induced coupling between liquid charge fluctuations and the internal excitations of the confining solid. Here, we develop a microscopic theory of exciton-mediated solid–liquid friction based on the coupling between optically generated excitons and charge fluctuations in water. We distinguish between static excitons, localized by disorder or functionalization, and dynamic excitons, which interact with water through polarization fluctuations. In both cases, we derive analytical formulas for the excitonic friction, which is experimentally tunable and can significantly reduce the slip length and thereby the hydraulic permeability of nanochannels. Applying our framework to carbon nanotubes, we quantitatively reproduce the recent measurements of Kistwal et al. (Nature 2026), showing a reduction of nanotube diffusion under optical excitation, without fitting parameters. More broadly, our results establish excitons as a mechanism to optically control nanofluidic transport and suggest that excitonic photoluminescence could provide an optical probe of flow velocity inside nanochannels.
\end{abstract}

\maketitle


The engineering of nanochannels exhibiting high flow rates is a central objective of nanofluidics, with direct implications for membrane-based technologies~\cite{Faucher2019, Emmerich2024, Tristan2020, Logan2012, Siria2017}.
At these scales, liquid transport is dominated by interfacial properties, as fluids exhibit finite slip at solid–liquid interfaces~\cite{Joly2006, Kavokine2021}.
This slip is governed by the interfacial friction coefficient,
typically set by surface roughness and chemical heterogeneity, 
and can become remarkably large on atomically smooth materials such as graphene and carbon nanotubes (CNTs), leading to flow enhancements of up to three orders of magnitude~\cite{Holt2006, Maali2008, Falk2010, Secchi2016, Tocci2020}.

The search for materials exhibiting large slip has recently led to the observation of an additional dissipation mechanism, first identified in non-contact solid–solid friction~\cite{Pendry1997, Persson1998, Volokitin1999, Volokitin2006}, arising from the coupling between liquid charge fluctuations and solid excitations, including electrons, plasmons, and phonons~\cite{Kavokine2022, Bui2023, Lizee2024}.
In water, the dominant charge fluctuations—associated with Debye relaxation and librational modes—span frequencies below $\sim \qty{20}{\tera\hertz}$ and behave as effective quasiparticles termed \emph{hydrons}, enabling coupling to low-frequency excitations in solids~\cite{Carlson2020, Kavokine2022, Coquinot2023b, Coquinot2023}.
This fluctuation-induced coupling establishes a direct connection between fluid dynamics and condensed matter physics, giving rise to transport phenomena such as electron drag~\cite{Narozhny2016, Lizee2023, Coquinot2023, Herrero2026}, band structure effects~\cite{Ambrosetti2022, Ambrosetti2023, ambrosettiBallisticAtomicTransport2026}, and quasiparticle tunneling~\cite{Coquinot2025}.
Consequently, liquid transport becomes intrinsically coupled to the solid’s internal degrees of freedom, so that the slip length is no longer an interfacial constant but is governed by the state of the solid excitations~\cite{Coquinot2024, Coquinot2026}.

Thus, controlling the state of the solid provides a route to engineer liquid transport. 
One experimental approach to tune this state is through the generation of excitons, the hydrogen-like bound states of a conduction-band electron and a positively charged hole~\cite{Frenkel1936Absorption, Gross1962Excitons}. 
Excitons are ubiquitous quasiparticles in semiconductors and can be generated optically: photon absorption creates electron–hole pairs that bind via Coulomb interaction.
They are not restricted to bulk materials but also emerge in low-dimensional semiconducting systems~\cite{Maultzsch2005ExcitonBinding, scharfExcitonicStarkEffect2016}.
In particular, excitons dominate the photophysics of single-walled CNTs~\cite{Korovyanko2004Ultrafast, kilinaCrosspolarizedExcitonsCarbon2008}.
Although excitons eventually recombine, emitting phonons or photons and thereby enabling their detection via photoluminescence experiments~\cite{Kira1998Microscopic, birkmeierProbingUltrafastDynamics2022}, their steady-state concentration can be efficiently controlled by the intensity of the incident light~\cite{LAUSSY2024706, Forno2024ExcitonLasing, Mishra2026ControllingQuasiparticlePopulation}.
The energy required to create excitons is large compared to thermal energy at room temperature~\cite{Maultzsch2005ExcitonBinding}. 
However, once optically generated, they provide low-frequency degrees of freedom that can respond to electrostatic interactions by polarization~\cite{scharfExcitonicStarkEffect2016}.

In particular, they may couple with charge fluctuations of a dipolar liquid. 
\response{This mechanism has recently been invoked in experiments probing solid–liquid friction in optically excited CNTs, supported by classical simulations~\cite{Kistwal2026, Kavokine2026}. However, no theoretical treatment accounting for the fundamentally quantum nature of excitons has yet been developed.
Indeed, unlike particles typically considered in quantum-friction problems~\cite{Kavokine2022, Bui2023, Lizee2024}, excitons possess an internal degree of freedom, the electron--hole distance, quantized into a Rydberg-like spectrum~\cite{lefebvreExcitedExcitonicStates2008}. Moreover, excitons in pristine nanotubes are neutral and carry no static dipole moment in any internal state~\cite{kilinaCrosspolarizedExcitonsCarbon2008}. 
Their coupling to fluctuating water charges therefore proceeds through discrete inter-state transitions, a feature not captured by existing descriptions of excitonic drag~\cite{bermanDragEffectsSystem2010a} and friction~\cite{Kistwal2026}.}

\begin{figure*}[t]
    \centering
    \includegraphics[width=1\linewidth]{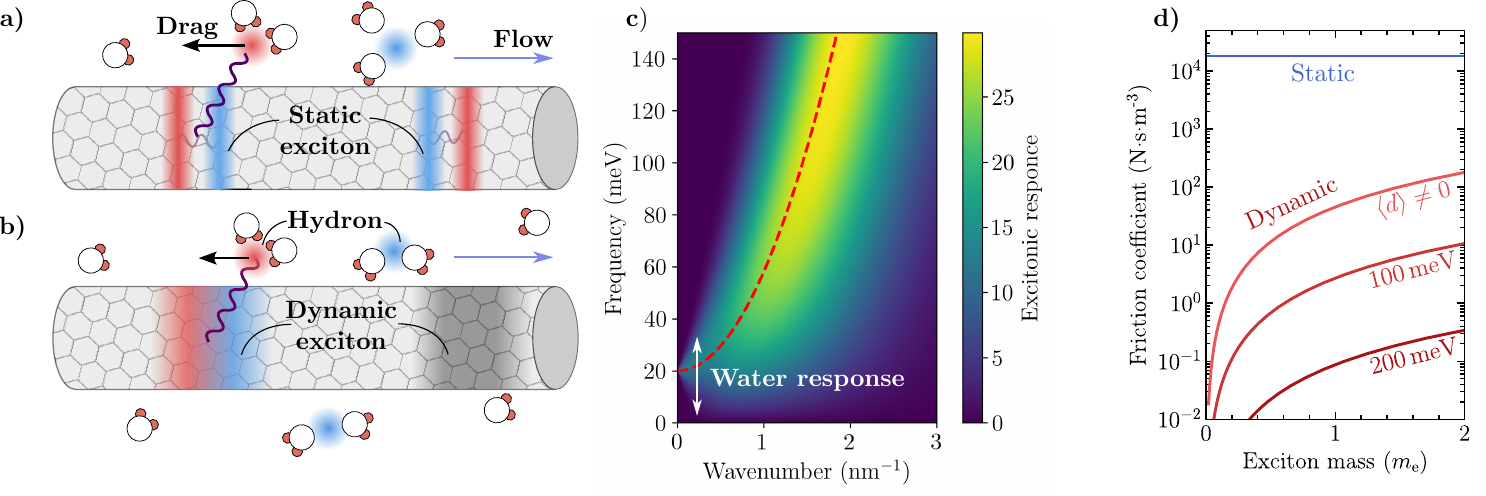}
    \caption{\textbf{a)}~Sketch of fluctuation-induced friction with static (pinned) excitons having a static dipole moment.
    \textbf{b)}~Sketch of fluctuation-induced friction with dynamic excitons that have zero static dipole moment (represented with gray), but can be polarized by the liquid-induced electric field. \textbf{c)}~Response function $g_{\mathrm{ex}}(q,\omega)$ of dynamic excitons. The red dashed line corresponds to transition from excitonic ground state, $\hbar\omega=\mathcal{E}^\alpha_q-\mathcal{E}^0_0$. The white arrow marks the region of dominant spectral weight of the water response function. Exciton and nanotube parameters are: $\Delta E=\qty{20}{\milli\eV}$, $M=m_{\mathrm{e}}$, $n_{\mathrm{ex}}^{\mathrm{1D}}=\qty{e6}{\per\centi\meter}$, $|d_{01}|=\qty{1}{\nano\meter}$, $a=\qty{0.39}{\nano\meter}$ (single-walled CNT with chirality (6,5)). \textbf{d)}~Friction coefficient of dynamic excitons as a function of their effective mass and for a transition energy from bottom to top $\Delta E=\qty{200}{\milli\eV}$, $\qty{100}{\milli\eV}$. \response{The highest red curve represents dynamic excitons with nonzero static dipole in the ground state}.
   The other parameters are the same than in panel (c). The blue line corresponds to static excitons with the same density.}
    \label{fig:1}
\end{figure*}

\response{In this Letter, we develop a microscopic quantum theory of fluctuation-induced friction between driven excitons and water. We derive analytical expressions for the excitonic friction, and investigate its consequence for nanotube hydrodynamic properties, including comparison with the experimental results of Ref.~\cite{Kistwal2026}.
As a first step, we consider a semiclassical model of static excitons, pinned at the interface and characterized by a permanent dipole moment, as illustrated in Fig.~\ref{fig:1}a.}
Such excitons typically appear in the presence of lattice defects and external potentials created by functionalizing molecules, such as DNA~\cite{zhengPhotoluminescenceDynamicsDefined2021}.
The electron and the hole then form localized states and act as an effective electrostatic roughness for the liquid. 
Second, we consider dynamic excitons, which are mobile along the interface and do not possess a permanent dipole moment.
Instead, they can be polarized by the fluctuating electric field generated by the liquid, leading to a dynamical coupling, as illustrated in Fig.~\ref{fig:1}b.
In both cases, we compute the exciton–water friction in CNTs for concreteness, although our framework is general and applies to any solid interface supporting excitons.
As a validation, we apply our model to the experimental conditions of~\citet{Kistwal2026} and obtain good agreement.
Overall, our results demonstrate that excitons provide a viable route to optically control liquid transport.

We first consider static excitons, described by localized electron–hole pairs pinned at the interface (see Fig.~\ref{fig:1}a). \response{In the following we assume that the nanotubes are very narrow, compared to the spatial extension of exciton's wavefunction (i.e., excitonic Bohr radius), which is typically several nanometers for Wannier--Mott excitons~\cite{Maultzsch2005ExcitonBinding, kilinaCrosspolarizedExcitonsCarbon2008}. }
\response{Because the exciton Bohr radius greatly exceeds the lattice spacing, we treat excitons as azimuthally symmetric one-dimensional objects~\cite{adamyan2008effects, hartmannExcitonsNarrowgapCarbon2011} within the effective-mass approximation~\cite{bastard1990wave}.}
Within the static model, an exciton is characterized by a linear charge density $e\Lambda(z)$ with zero net charge. The charge distribution $\Lambda$ generates an electric potential $V_\Lambda(z,\rho)$ (where $\rho$ denotes the radial distance from the nanotube) which polarizes the surrounding water both inside and outside the tube.
In turn, the polarized liquid produces an induced potential $V_{\rm ind}(z,\rho)$ that interacts with $\Lambda$. 
The strength of this response is encoded in the (retarded) surface response function $g_\mathrm{w}^{\mathrm{R}}$ defined by $V_{\mathrm{ind}}(q, \rho=a, \omega)=-g_\mathrm{w}^{\mathrm{R}}(q, \omega)V_\Lambda(q, \rho=a, \omega)$ where $q$ is the wavevector along the tube axis and the frequency~$\omega$ encodes the finite-time polarization response of water.
This response function is directly related to the charge susceptibility (or equivalently the dielectric function) of water and in general depends on the geometry (see SM~Sec.~1.2)~\cite{Coquinot2023b, Gispert2025}. 
However, since friction is dominated by wavenumbers $q\gtrsim 1/a$, for which the interface appears locally flat, we approximate $g_\mathrm{w}$ by its planar expression~\cite{Kavokine2022}. 
When water flows with velocity $v_\mathrm{w}$, the induced electric potential $V_{\rm ind}$ lags behind the excitonic charge distribution $\Lambda$, resulting in a net electrostatic force from their interaction.
In Fourier space, this delay is captured by the Doppler shift $\omega\rightarrow \omega-qv_\mathrm{w}$ in the surface response function of water.
To linear order in $v_\mathrm{w}$, the total friction force reads $F=-\mathcal{A}\lambda_{\rm ex}^{\rm stat}v_\mathrm{w}$ where $\mathcal{A}$ is the nanotube surface area. 
The corresponding excitonic friction coefficient is (see SM~Sec.~1):
\begin{equation}\label{eq:lambda_trapped}
\lambda_{\mathrm{ex}}^{\mathrm{stat}}=\frac{e^2}{4\pi\epsilon_0}\frac{n_{\mathrm{ex}}^{\mathrm{1D}}}{\pi a^2}\int_0^\infty \frac{dq}{2\pi}\; q\;|\Lambda(q)|^2\frac{\partial}{\partial \omega}\mathrm{Im}\left[g_{\mathrm{w}}^\mathrm{R}(q,0)\right],\end{equation}
where $n_{\mathrm{ex}}^{\mathrm{1D}}$ is the linear density of excitons along the nanotube and $\epsilon_0$ is the vacuum permittivity. 
For excitons larger than the liquid charge correlation length (\AA-scale), the liquid probes a local charge and we can approximate $|\Lambda(q)|^2 \approx 2$ if electron and hole are strongly localized. For a linear excitonic density of ${\qty{e6}{\per\centi\meter}}$, we then predict an excitonic friction of $\lambda_{\rm ex}^{\rm stat}\approx \qty{e4}{\newton\second\per\meter\cubed}$ (see Fig.~\ref{fig:1}d), comparable to roughness-induced friction for flows outside CNTs and exceeding that of flows inside narrow CNTs by one to two orders of magnitude~\cite{falkMolecularOriginFast2010a}. 

We now turn to dynamic excitons, modeled as one-dimensional free particles with an internal degree of freedom corresponding to the relative electron–hole coordinate ${\hat{d}=\hat{z}_{\mathrm{h}}-\hat{z}_{\mathrm{e}}}$ (see Fig.~\ref{fig:1}b). 
This operator represents the excitonic dipole moment. 
Its spectrum is quantized, with a symmetric ground state exhibiting no static dipole, $\braket{0|\hat{d}|0}=0$, while finite transitional dipole moment exists between ground and excited states~\cite{lefebvreExcitedExcitonicStates2008}. 
At equilibrium, excitons predominantly occupy the ground state, although a finite population of excited states arises from thermal activation.
When coupled to water, the exciton experiences a fluctuating electric field that can induce transitions from the ground state to excited states, thereby polarizing the exciton.
Similarly to water, this response is captured by an excitonic surface response function $g_{\mathrm{ex}}^{\mathrm{R}}$, which can be computed analytically from the dipole operator and the transitions between excitonic states (see SM~Sec.~2.3):
\begin{equation}\label{eq:g_ex}
g_{\mathrm{ex}}^{\mathrm{R}}(q,\omega)\!\!=\!\! -\frac{e^2}{4\pi\epsilon_0 a}\!\sum_{\alpha,\beta}\!\int_{\bar{q}}\!\frac{d\bar{q}}{2\pi}\frac{|{d}^{\alpha\beta}|^2\!\left[ n\left(\mathcal{E}^{\alpha}_{\bar{q}}\right)\!\!-\!n\left(\mathcal{E}^{\beta}_{q+\bar{q}}\right)\!\right] }{\hbar\omega+\mathcal{E}^{\alpha}_{\bar{q}}-\mathcal{E}^{\beta}_{q+\bar{q}}+i0},
\end{equation}
where $n(\epsilon)=n_{\mathrm{ex}}^{\rm 1D} \exp[-\epsilon/k_BT]/\mathcal{Z}$ is the Boltzmann distribution with partition function $\mathcal{Z}$ and temperature $T$. ${\mathcal{E}^{\alpha}_{q} = E_\alpha + \hbar^2q^2/2M}$ denotes the energy of a state combining an internal electron-hole state $\alpha$ of energy $E_\alpha$, and center-of-mass motion with momentum $\hbar q$ and effective mass $M$. These results may be easily extended to excitons in 2D semiconducting materials (see SM~Sec.~2.5). The resulting response function for nanotube excitons is shown in Fig.~\ref{fig:1}c at room temperature.
The dispersion relation associated with transitions from the ground state (in both dipole and center-of-mass degrees of freedom), which dominates at low temperature, is shown as a red dashed line.

\begin{figure*}[th]
    \centering
    \includegraphics[width=1\linewidth]{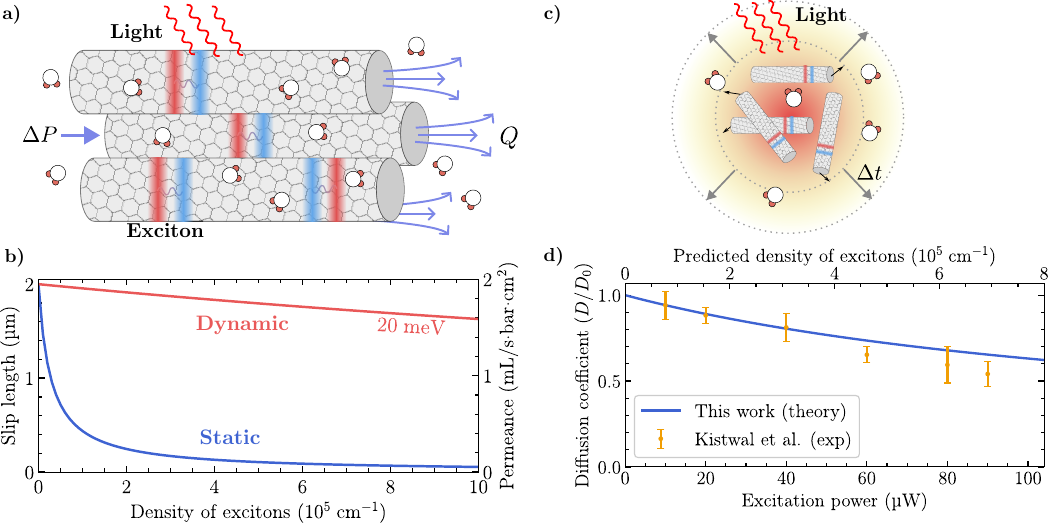}
    \caption{\textbf{a)}~Sketch of the exciton-mediated permeance through a membrane made of nanotubes. \textbf{b)}~Total slip length and membrane permeance (assuming 50\% packing fraction) as a function of linear density of excitons. Dynamic excitons (red) have $|d^{01}|=\qty{1}{\nano\meter}$, $M=2m_{\mathrm{e}}$, $\Delta E=\qty{20}{\milli\eV}$. Nanotube radius and length are $a=\qty{0.39}{\nano\meter}$ and $L=\qty{1}{\micro\meter}$. Non-excitonic slip length inside the nanotubes is $b_0=\qty{2}{\micro\meter}$. \textbf{c)}~Sketch of the experiment of~\citet{Kistwal2026}:  CNT diffusion is measured while a laser creates excitons. \textbf{d)}~Nanotube diffusion coefficient $D$ normalized by the exciton-free diffusion coefficient $D_0$ as a function of the applied excitation power $P$ from our analytical model and from the experimental measurements of~\citet{Kistwal2026}.
 Model values: $\sigma=\qty{e-17}{\centi\meter\squared}$~\cite{berciaudLuminescenceDecayAbsorption2008}, $\tau_{\rm ex}=\qty{1}{\nano\second}$~\cite{heIntrinsicLimitsDefectstate2019}, $b_0=\qty{50}{\nano\meter}$~\cite{falkMolecularOriginFast2010a}. From~\citet{Kistwal2026} we used $A_{\mathrm{cf}}=\qty{0.28}{\micro\meter\squared}$, $E_{\mathrm{ph}}=\qty{2.6}{\eV}$. The top axis is the predicted density of excitons $n_{\mathrm{ex}}^{\mathrm{1D}}$ according to our model.
 }
    \label{fig:2}
\end{figure*}

In contrast to static excitons, dynamic excitons exhibit no net electrostatic dipole on average.
However, temporal fluctuations of the dipole moment generate transient charge distributions, which give rise to dissipation.
Summing over these fluctuations across frequencies results in a net friction force.
Following~\cite{Kavokine2022}, the resulting fluctuation-induced friction can be computed within perturbation theory. 
To leading order in both the water velocity and the interaction, the excitonic friction coefficient reads (see SM~Sec.~2):
\begin{equation}\label{eq:lambda_quantum}
    \lambda_{\rm ex}^{\rm dyn}\!=\!\frac{\hbar^2}{4\pi^3 k_\mathrm{B}T a}\!\int_0^\infty \!\!\!\!\!dq\; q^2 \!\!\int_0^\infty \!\!\!\!\!\! d\omega \frac{\mathrm{Im}\!\left[g_{\mathrm{ex}}^{\mathrm{R}}(q,\omega)\right]\mathrm{Im}\!\left[g_{\mathrm{w}}^{\mathrm{R}}(q,\omega)\right]}{\sinh^2(\hbar\omega/2k_\mathrm{B}T)},
\end{equation}
Friction arises from the spectral overlap between the excitonic and water surface response functions at frequencies populated by thermal fluctuations.
Consequently, if the energy gap to the first dipolar excitation exceeds the energy range of the water surface response function (below $\qty{30}{\milli\eV}$, see Fig.~\ref{fig:1}c), the overlap is reduced and excitonic friction is suppressed.
The resulting friction coefficient as a function of the exciton mass is shown in Fig.~\ref{fig:1}d (for a density~$n_{\mathrm{ex}}^{\mathrm{1D}}=\qty{e6}{\per\centi\meter}$) and for different excitation gaps, \response{including a model of a dynamic exciton with finite dipole moment in the ground state}.
Heavier excitons have a flatter band and couple more efficiently to water fluctuations, as they can accommodate large momenta at a lower kinetic energy cost.
Accordingly, we find the excitonic friction to scale with $\lambda_{\rm ex}^{\rm dyn}\propto M^{2}$ when $M\sim m_{\mathrm{e}}$. A similar scaling is obtained also for excitons in 2D materials.
For small excitation gaps and relatively high exciton masses, we predict an excitonic friction of $\lambda_{\rm ex}^{\rm dyn}\approx \qty{e2}{\newton\second\per\meter\cubed}$ in CNTs.

In all cases, we find that excitonic friction is larger for static excitons than for dynamic ones.
This originates from the fact that the water response is maximal at low frequencies, and therefore couples most efficiently to static disorder or to low-frequency solid excitations.
\response{In fact, static excitons correspond to the infinite-mass, polarized-ground-state limit, for which Eqs.~\eqref{eq:g_ex}--\eqref{eq:lambda_quantum} reduce to Eq.~\eqref{eq:lambda_trapped} (SM Sec.~2.4). Visually, the highest red curve in Fig.~\ref{fig:1}d approaches the blue curve at large masses. The static friction also rapidly saturates with dipole moment (SM Sec.~1.1), leaving no tunable excitonic parameter. Such a limit can arise from external potentials or disorder~\cite{singh2017ExcitonLocalization}, suggesting a stronger excitonic friction at defective or chemically functionalized surfaces.}
Nevertheless, even smooth surfaces hosting mobile excitons may display measurable signatures of this mechanism, provided the excitons have sufficiently large effective mass and a small excitation gap.

In particular, if exciton generation significantly enhances the solid–liquid friction, it correspondingly reduces the slip length. Specifically, it adds friction $\lambda_{\mathrm{ex}}$ to the non-excitonic friction $\lambda_0$, and the slip length in the presence of excitons then reads
\begin{equation}\label{eq:b}
    b\left(n_{\mathrm{ex}}^{\mathrm{1D}}\right) = \frac{b_0}{1+\lambda_{\mathrm{ex}} b_0/\eta},
\end{equation}
where $\eta$ is the viscosity and $b_0=\eta/\lambda_0$ is the bare slip length in the absence of excitons.
In confined CNTs, $b_0$ can reach micrometer scales, corresponding to small friction coefficients $\lambda_0\sim\qty{5e2}{\newton\second\per\meter\cubed}$~\cite{falkMolecularOriginFast2010a, Secchi2016}.
The resulting flow rate is enhanced by the slip according to the modified Hagen–Poiseuille law~\cite{kavokineFluidsNanoscaleContinuum2021}:
\begin{equation}\label{eq:flow_rate_large_slip}
    Q=\frac{\pi a^4 \Delta P}{8\eta L}\left(1+\frac{4b}{a}\right)\approx\frac{\pi a^3 \Delta P}{2\eta L}b\left(n_{\mathrm{ex}}^{\mathrm{1D}}\right),
\end{equation}
where $L$ is the nanotube length and $\Delta P$ the applied pressure drop. 
This formula can further be extrapolated to estimate the permeance of a membrane (flow rate per pressure drop and area). 
The predicted slip length and permeance as functions of the exciton density are shown in Fig.~\ref{fig:2}b, exhibiting a marked reduction of flow in the presence of excitons. The effect is more significant for the case of static excitons because they exhibit stronger friction; however, the effect is noticeable for dynamic excitons too thanks to the low roughness inside CNTs. Thus, exciton generation provides a  route to optically control the slip length and, consequently, the permeance of nanochannels and membranes.

Finally, we compare our theory to the recent experiment of \citet{Kistwal2026}, in which the diffusion coefficient of individual CNTs, measured via fluorescence correlation spectroscopy (FCS), was observed to decrease with pumping power, which was attributed to excitonic friction (see Fig.~\ref{fig:2}c). 
To analyze these observations within our framework, we first relate the diffusion coefficient of a CNT to the solid–liquid friction.
When moving, a nanotube experiences anisotropic hydrodynamic drag tensor $\boldsymbol{\xi}$, with $\xi_\parallel$ along its axis—primarily set by the solid–liquid friction outside the tube—and $\xi_\perp$ normal to it.
These drag coefficients can be approximated within slender-body theory (SBT) by solving the Stokes equations locally along the cylinder in the large aspect ratio limit $L\gg a$~\cite{batchelorSlenderbodyTheoryParticles1970, bloomfieldFrictionalCoefficientsMultisubunit1967c, kellerSlenderbodyTheorySlow1976a, tiradoTranslationalFrictionCoefficients1979}.
This approximation is well justified for the single-walled CNTs used in~\citet{Kistwal2026}, where $L\approx\qty{600}{\nano\meter}$ and $a\approx\qty{0.4}{\nano\meter}$.
Averaging the drag tensor over slow rotational diffusion, the Einstein--Smoluchowski relation gives the diffusion coefficient (see SM~Sec.~3):
\begin{equation}\label{eq:CNT_diffusion_equation}
D=\frac{k_{\mathrm{B}}T}{3}\mathrm{Tr}(\boldsymbol{\xi}^{-1})=\frac{k_{\mathrm{B}}T}{3\pi\eta L}\left[\log\left(\frac{L}{2a}\right)+\gamma+\frac{b}{2a}\right].
\end{equation}
Here, $\gamma$ is a small, bounded correction to the slender-body limit~\cite{tiradoTranslationalFrictionCoefficients1979, aragonHighPrecisionTransport2009}, negligible in our situation. Eq.~\eqref{eq:CNT_diffusion_equation} extends previous models for CNT diffusion~\cite{waltherHydrodynamicPropertiesCarbon2004, nairDynamicsSurfactantSuspendedSingleWalled2008, streitMeasuringSingleWalledCarbon2012, rudyakDiffusionSinglewalledCarbon2020, leeLengthMeasurementSinglewalled2020} by incorporating a finite slip length $b$ at the outer solid–liquid interface, which significantly reduces the axial drag $\xi_\parallel$.
In the no-slip regime $b\ll a$, the diffusion coefficient predicted by Eq.~\eqref{eq:CNT_diffusion_equation}, $D_{0} \approx \qty{4.9}{\micro\meter\squared\per\second}$, already exceeds the experimental measurements of~\citet{Kistwal2026}, which reach $D_{0}^{\rm exp} \approx \qty{0.9}{\micro\meter\squared\per\second}$. 
The discrepancy becomes even larger in the presence of slippage.
Several effects may explain why the theoretical model overestimates the measured absolute values \response{(See SM Sec. 3.6)}. First, the CNTs exhibit a broad length distribution~\cite{streitMeasuringSingleWalledCarbon2012}; since longer nanotubes absorb and emit more light, they dominate the FCS signal and bias the inferred diffusion coefficient toward smaller values. Second, the effective hydrodynamic radius likely exceeds the bare CNT radius, in particular due to the DNA wrapping~\cite{campbellAtomicForceMicroscopy2008, kodeInteractionDNAComplexedBoron2021}. Third, the nanotube concentration (\qty{1}{\nano\molar}) corresponds to an average inter-tube spacing of order \(\qty{1}{\micro\meter}\), comparable to the CNT length, which may induce steric effects. \response{Importantly, these effects are distinct from excitonic friction, which only controls the slip length. Such discrepancies are well documented and commonly accounted for by a phenomenological prefactor in Eq.~\eqref{eq:CNT_diffusion_equation}~\cite{tsyboulskiTranslationalRotationalDynamics2008, streitMeasuringSingleWalledCarbon2012}.}
In the large-slip regime $b\gg a$, transport is dominated by axial motion and the diffusion coefficient scales linearly with $b$. 
\response{In the absence of excitons, the outer slip length of CNTs was estimated to be \(b_0\approx 50\) nm~\cite{falkMolecularOriginFast2010a}, which we will use despite the DNA wrapping.}
Exciton generation then increases the solid–liquid friction, reducing the slip length via Eq.~\eqref{eq:b} and, consequently, the diffusion coefficient.

Experimentally, a laser of power \(P\), corresponding to a photon flux \(P/E_{\mathrm{ph}}\) (with \(E_{\mathrm{ph}}\) the photon energy at the laser wavelength), is focused onto a confocal area \(A_{\mathrm{cf}}\). The laser energy is tuned to resonantly create excitons in the nanotubes. Assuming that each absorbed photon generates one exciton, and denoting by \(\sigma\) the absorption cross section per unit nanotube length, the steady-state exciton density is obtained by balancing the exciton generation rate against exciton decay into phonons and photons, characterized by a lifetime \(\tau_{\mathrm{ex}}\). This yields
\begin{equation}
n_{\mathrm{ex}}^{\rm 1D}= \frac{P \sigma \tau_{\mathrm{ex}}}{A_{\mathrm{cf}} E_{\mathrm{ph}}}.
\end{equation}
\response{Because the experiment uses DNA-functionalized CNTs, which can localize excitons~\cite{zhengPhotoluminescenceDynamicsDefined2021} and induce a permanent ground-state dipole (SM Sec.~2.6), we model the excitons as effectively static and compute the friction versus power $P$ using Eq.~\eqref{eq:lambda_trapped}.} Using Eqs.~\eqref{eq:b}–\eqref{eq:CNT_diffusion_equation} with parameter values extracted from the experiment and the literature (SM~Sec.~3.5), we predict a reduction of the diffusion coefficient due to excitonic friction, as shown in Fig.~\ref{fig:2}d.
Focusing on relative variations, we find good quantitative agreement with the experimental results of~\citet{Kistwal2026}. 
For instance, for a laser power of $\qty{100}{\micro\watt}$, we predict a reduction of the diffusion coefficient of around 40\%, consistent with the experiment.
Our results thus support the interpretation of the experimentally observed effect as an exciton-driven increase in friction. 


In conclusion, we have developed a microscopic theory of exciton-mediated solid–liquid friction, showing that excitons generate a significant additional dissipation channel for the liquid. 
\response{Applying a slender-body hydrodynamic description with finite slip, we showed that this framework quantitatively captures the recently observed reduction of nanotube diffusion under optical excitation~\cite{Kistwal2026, Kavokine2026}. This experiment therefore provides evidence for excitonic friction and its optical control, while our theory enables predictions beyond this specific configuration.}
We find the excitonic friction to be particularly large for static excitons, which are typically localized by functionalizing molecules, but it could also be relevant for dynamic excitons in nanotubes or 2D materials having a large effective mass and a small polarization energy barrier. In particular, transition-metal dichalcogenides such as MoSe\textsubscript{2} monolayers, which exhibit large effective masses and excitation energies of order $\sim\qty{5}{\milli\eV}$, are promising materials for strong friction induced by dynamic excitons~\cite{liuExcitonpolaronRydbergStates2021}. Overall, our results show that excitons provide an experimentally accessible route to optically control liquid transport through nanochannels. 

Moreover, through excitonic friction, a hydrodynamic flow could bias the momentum distribution of dynamic excitons in favor of motion along the flow, analogous to Coulomb drag~\cite{Narozhny2016, Coquinot2023}. Excitonic photoluminescence could thereby enable optical measurement of the liquid velocity inside a nanochannel. Exciton-mediated friction thus opens a new direction at the interface between condensed matter physics and nanofluidics, where optical excitation becomes a tool to control and probe nanoscale flows.

\section*{Acknowledgements} 
The Authors thank Nikita Kavokine, Marialore Sulpizi and Sebastian Kruss for fruitful discussions. B.C. acknowledges support from the NOMIS Foundation.

\bibliographystyle{apsrev4-2}
\bibliography{bibliography, bibfile}

\end{document}


\maketitle
\section*{Conventions}
We define the Fourier transform and its inverse as:
\begin{equation}
    f(q,\omega)=\int dz \;dt\; f(z,t)\;e^{-iqz+i\omega t},\quad f(z,t)=\int \frac{dq}{2\pi} \;\frac{d\omega}{2\pi}\; f(q,\omega)\;e^{iqz-i\omega t}
\end{equation}
\section{Static excitons}

We assume that static excitons are pinned to the nanotube and the electron and hole wavefunction are radially symmetric. The linear charge density of an exciton is given by $e\Lambda(z)$, and by electric neutrality, $\int\Lambda(z)dz=0$. 

\subsection{Friction coefficient for a nanotube}

We work in the reference frame of water in which excitons are moving steadily along the nanotube. They exert a variable electric field, which polarized the fluid, as measured by its response function, and we can calculate the force exerted on the polarized charge by the excitons. Assume first that the water flows outside of the nanotube (the calculations are very similar for the case of internal flow). Assuming the excitons are confined to the nanotube surface, $r=a$, the potential (in units of energy) exerted by an exciton at a given point $(r,\phi,z)$ in cylindrical polar coordinates is given by
\begin{equation}
    \varphi(r,z, t)=\int dz' \; \Lambda(z') V^{\mathrm{C}}(z+vt-z', r, a)
\end{equation}
where $V^{\mathrm{C}}$ is the radially symmetric component of the Coulomb potential expressed in cylindrical coordinates. Its Fourier transform is
\begin{equation}
    V^{\mathrm{C}}(q; r,a)=\frac{e^2}{4\pi\varepsilon_0}\,2I_0(qa)K_0(qr),
\end{equation}
where $I_0$, $K_0$ are the modified Bessel functions of order 0 for which we use convention $I_0(-x)=I_0(x)$, $K_0(-x)=K_0(x)$. The potential inside the nanotube is given by an analogous expression with a change $I_0\leftrightarrow K_0$. The potential $\varphi$ induces water charge density given by 
\begin{equation}
    n_{\mathrm{w}}(\mathbf{x},t)=\int_{-\infty}^\infty dt' \int_V d\mathbf{x}' \chi_{\mathrm{w}}^{\mathrm{R}}(\mathbf{x},\mathbf{x}',t,t')\varphi(\mathbf{x'},t'),
\end{equation}
where the integration is over the entire volume $V$ occupied by the fluid, which is $\{r>a\}$ or $\{r<a\}$, and $\chi_\mathrm{w}^\mathrm{R}$ is the retarded susceptibility function, which is zero for $t<t'$. The water charge creates an induced potential, which exerts a nonzero force on the excitons:
\begin{equation}
    F_z=\int_{-\infty}^\infty dz \Lambda(z)\left(-\frac{\partial}{\partial z}\right)\int d\mathbf{x}' n_{\mathrm{w}}(\mathbf{x'},t)V^{\mathrm{C}}(z-z', r',a)
\end{equation}

Performing the Fourier transform, we arrive at 
\begin{equation}\label{eq:supplementary_force_z_static}
\begin{aligned}
    F_z&=\frac{e^2}{4\pi \epsilon_0}\int\frac{dq}{2\pi} (iq) |\Lambda(q)|^22I_0(qa)K_0(qa)\\
    &\times\left[-\frac{e^2}{4\pi \epsilon_0}\int dA' dA'' \frac{4 I_0(qa)^2 K_0(qr')K_0(qr'')}{2I_0(qa)K_0(qa)}\chi_{\mathrm{w}}^{\mathrm{R}}(\mathbf{r}',\mathbf{r}'', q, \omega=qv)\right].
    \end{aligned}
\end{equation}
Here integration over $dA'=r'dr'd\phi'$ represents integration in the plane normal to the nanotube axis for $r'>a$. The expression in the square brackets is the appropriate definition of the radially symmetric confined water response function $g_{\mathrm{w}}^{\mathrm{R}}(q,\omega=qv)$ for the liquid outside the nanotube. This definition is consistent with the definition
\begin{equation}
    g_{\mathrm{w}}(z,t)=-\frac{\varphi_{\mathrm{ind}}(a,t)}{\varphi_{\mathrm{ext}}(a,t)}
\end{equation}
for a radially symmetric potential $\varphi_{\mathrm{ext}}$ induced by charges inside or on the surface of the nanotube.

The real part of $g_\mathrm{w}^\mathrm{R}(q,qv)$ is an even function of $q$, and correspondingly it does not contribute to the integral. Overall, using the odd parity of the imaginary part of the response,
\begin{equation}
    F_z=\frac{e^2}{4\pi\epsilon_0}\;2\int_{0}^\infty\frac{dq}{2\pi} q|\Lambda(q)|^2 2 I_0(qa)K_0(qa)\mathrm{Im}\,g_{\mathrm{w}}^{\mathrm{R}}(q,qv).
\end{equation}

To extract the friction coefficient, we calculate the force per unit surface area of the nanotube per unit relative velocity $v$. For that, we expand $\mathrm{Im}\,g_{\mathrm{w}}^{\mathrm{R}}$ to leading, linear order in velocity and assume a linear density of excitons $n_{\mathrm{ex}}^{\rm 1D}$. The friction coefficient is then
\begin{equation}\label{eq:supplementary_lambda_stat}
    \begin{aligned}
        \lambda_{\mathrm{ex}}^{\mathrm{stat}}&=\frac{n_{\mathrm{ex}}^{\rm 1D}}{2\pi a}\frac{\partial F_z}
        {\partial v}\Biggr|_{v=0}\\
        &=\frac{e^2}{4\pi\epsilon_0}\frac{n_{\mathrm{ex}}^{\rm 1D}}{\pi a}\int_0^\infty \frac{dq}{2\pi} q^2|\Lambda(q)|^2\; 2 I_0(qa)K_0(qa)\frac{\partial}{\partial \omega}\mathrm{Im}\,g_{\mathrm{w}}^\mathrm{R}(q,0).
    \end{aligned}
\end{equation}

\begin{figure}[ht]
    \centering
    \captionsetup{justification=centering}
    \includegraphics[width=0.7\linewidth]{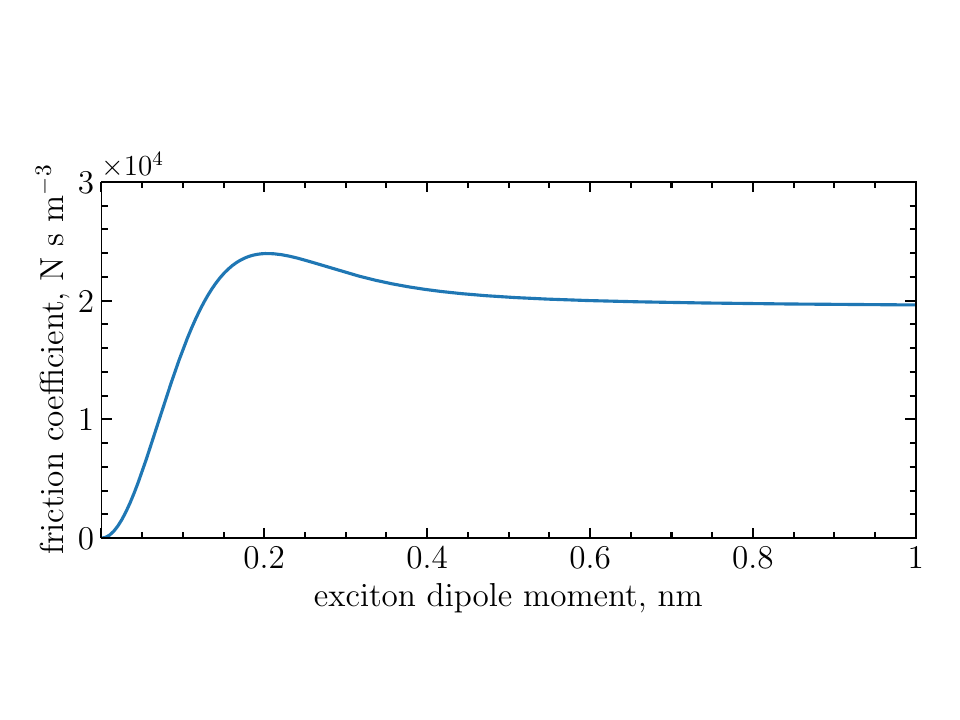}
    \caption{\response{Static friction coefficient $\lambda$ vs exciton dipole moment $d$ for $n_{\mathrm{1D}}=\qty{0.1}{\per\nano\meter}$, $a=\qty{0.37}{\nano\meter}$.}}
    \label{fig:lambda_vs_d}
\end{figure}

Assuming a static exciton charge distribution as a pair of opposite charged rings separated with distance $d$, the corresponding momentum-space distribution is $|\Lambda(q)|=2\sin(qd/2)$. The integral in Eq.~\eqref{eq:supplementary_lambda_stat} has typical wavenumbers of $q\sim \qty{1}{\angstrom}$, so the factor $|\Lambda(q)|^2$ is a quickly-oscillating function which can be replaced with its average value of $\langle |\Lambda(q)|^2\rangle=2$ in the limit of large $d$, corresponding to two localized particles. \response{This saturation is demonstrated by plotting friction coefficient as a function of dipole moment in Fig.~\ref{fig:lambda_vs_d}. From this Figure, we see that for $d\gtrsim \qty{0.1}{\nano\meter}$ the dependence of friction on the dipole moment is very weak. Therefore, existence of a static dipole moment makes the prediction of friction coefficient independent of the exact charge distribution, i.e., dipole moment.}

\subsection{Planar approximation}
The water response function varies over typical wavevector $q\sim\qty{10}{\per\nano\meter}$, which implies that $qa\gg 1$ for typical nanotube radius $a\sim\qty{1}{\nano\meter}$. Consequently, we may make the approximation~\cite{NIST:DLMF}:
\begin{equation}
    2 I_0(qa)K_0(qa)\approx 2\times\frac{1}{\sqrt{2\pi qa}}e^{qa}\times\sqrt{\frac{\pi}{2qa}}e^{-qa}=\frac{1}{qa}.
\end{equation}
Thus, the friction coefficient can be written as
\begin{equation}
    \lambda_{\mathrm{ex}}^{\mathrm{stat}}=\frac{e^2}{4\pi\epsilon_0}\frac{n_{\mathrm{ex}}^{\rm 1D}}{\pi a^2}\int_0^\infty \frac{dq}{2\pi} q|\Lambda(q)|^2\;\frac{\partial}{\partial \omega}\mathrm{Im}\,g_{\mathrm{w}}^\mathrm{R}(q,0),
\end{equation}
which is reported in the main text.

In this limit, the response function of water matches with the planar case. Indeed, consider the factor appearing in the definition of water response function:
\begin{equation}
    \begin{aligned}
        \frac{4 I_0(qa)^2 K_0(qr')K_0(qr'')}{2I_0(qa)K_0(qa)}&\approx 2\times \frac{1}{\sqrt{2\pi qa}}e^{qa}\times\sqrt{\frac{\pi}{2qr'}}e^{-qr'}\times \sqrt{\frac{\pi}{2qr''}}e^{-qr''}\times \sqrt{\frac{2q a}{\pi}} e^{qa}\\
        &=\frac{1}{q\sqrt{r'r''}} e^{-q(r'-a)-q(r''-a)}.
    \end{aligned}
\end{equation}
For $qa\gg1$ the exponential factor decays on the scale $\Delta r\sim a$ very quickly, so the relevant radii $r'$ and $r''$ are very close to the nanotube radius $a$, which allows us to replace the denominator with $q\sqrt{r'r''}\approx qa$. With this approximation, the water response function becomes
\begin{equation}
\begin{aligned}
    g_{\mathrm{w}}^{\mathrm{R}}(q,\omega)&=-\frac{e^2}{4\pi \epsilon_0}\int dA' dA'' \frac{1}{qa}e^{-q(r'-a)-q(r''-a)}\chi_{\mathrm{w}}^{\mathrm{R}}(\mathbf{r}',\mathbf{r}'', q, \omega)\\
    &=-\frac{e^2}{4\pi \epsilon_0} \int_a^\infty\!\!\!\!\int_a^\infty \!\!\!r'r'' dr'dr'' \!\!\int_0^{2\pi}\!\!\!\int_0^{2\pi} d\phi' d\phi''\frac{1}{qa}e^{-q(r'-a)-q(r''-a)}\chi_{\mathrm{w}}^{\mathrm{R}}(\mathbf{r}',\mathbf{r}'', q, \omega).
\end{aligned}
\end{equation}
As before, we may approximate $r'r''\approx a^2$ and for this section only, we can change the notation to the local Cartesian coordinates:
\begin{equation}
    (z, a\phi, r-a)\mapsto (x,y,z),
\end{equation}
so that the wavevector $q$ comes from the $x$-Fourier transform, so we denote it as $q_x$. Then
\begin{equation}
\begin{aligned}
    g_{\mathrm{w}}^{\mathrm{R}}(q,\omega)&=-\frac{e^2}{4\pi \epsilon_0} \int_0^\infty\!\!\!\!\int_0^\infty \!\!\! dz'dz'' \!\!\int_0^{2\pi a}\!\!\!\int_0^{2\pi a} dy' dy''\frac{1}{qa}e^{-q_x(z'+z'')}\chi_{\mathrm{w}}^{\mathrm{R}}(z', y',z'', y'', q_x, \omega).
\end{aligned}
\end{equation}
By translational symmetry, $\chi_{\mathrm{w}}$ depends only on $y'-y''$. In fact, this follows from the rotational symmetry of the nanotube in the original variables. Then, the double integral over $dy' dy''$ can be replaced with a single integral over $2\pi a\; dy$, which gives a $q_y=0$ Fourier component. If we let $\mathbf{q}=(q_x,q_y=0)$, we obtain
\begin{equation}
    g_{\mathrm{w}}^{\mathrm{R}}(q,\omega)=-\frac{e^2}{4\pi \epsilon_0} \int_0^\infty\!\!\!\!\int_0^\infty \!\!\! dz'dz'' \frac{2\pi}{q}e^{-q(z'+z'')}\chi_{\mathrm{w}}^{\mathrm{R}}(z',z'', \mathbf{q}, \omega),
\end{equation}
which matches with the definition of the response function in planar case~\cite{kavokineFluctuationinducedQuantumFriction2022}. This confirms that our definitions are consistent with those used previously in the literature. Moreover, it justifies the approximation of using the planar response function $g_{\mathrm{w}}^{\mathrm{R}}(q,\omega)$ even in narrow nanotubes. Indeed, the water response function peaks at large wavenumbers corresponding to short lengthscales $\sim \qty{1}{\angstrom}$, which are smaller than nanotube radius and circumference. We have consistently employed the planar approximation uniformly for calculations and figures of the main text, although we derive the equations in a general form here.

\section{Dynamic excitons}

We now move to the derivation of friction with dynamic excitons. For that, we characterize the internal structure of an exciton by a relative position operator $\mathbf{d}=\mathbf{x}_e-\mathbf{x}_h$, where indices $e$ and $h$ refer to an electron and a hole, respectively. In general, this operator has three components $d_i$ with $i=x,\,y,\,z$. Its matrix elements are defined by
\begin{equation}
    d_i^{\alpha\beta}=\braket{\alpha|d_i|\beta}.
\end{equation}
The elements characterized by $\alpha=\beta$ describe the static dipole moment, while elements with $\alpha\neq \beta$ correspond to the transitional dipole moments.

If the wavefunctions of the excitonic states $\psi^{\text{rel}}_\alpha(\mathbf{d})$ are known, the corresponding matrix elements can be written as
\begin{equation}
    d_i^{\alpha\beta}=\int \mathrm{d}\mathbf{x}\; \bar{\psi}^{\text{rel}}_\alpha(\mathbf{x}) \; x_i\;\psi^{\text{rel}}_\beta(\mathbf{x}).
\end{equation}

\subsection{Keldysh charge correlation function for a nanotube}

We restrict ourselves to the case of a one-dimensional model of excitons with $\mathbf{d}$ pointing in the $z$-direction for considered intraband transitions~\cite{kilinaCrosspolarizedExcitonsCarbon2008} and denote $d_{z}=d$.

The exciton as a whole can be described by a one-dimensional wavefunction with an internal state index: $\psi_\alpha(z,t)$. It can be decomposed into a modes characterized with different wavenumbers $q$. Assuming the mass of the exciton $M$ remains constant, we have a dispersion relation of the form
\begin{equation}\label{eq:supplementary_exciton_dispersion_supplementary}
    \mathcal{E}^\alpha_q=E_{\alpha}+\frac{\hbar^2q^2}{2M},
\end{equation}
where $E_\alpha$ is the internal energy of the corresponding state.

The exciton-hydron interaction Hamiltonian is modeled as a charge-dipole electrostatic interaction:
\begin{equation}
    \begin{aligned}
    \mathcal{H}_{\mathrm{int}}&=\int dz_{\mathrm{ex}} \int d{\mathbf{x}}_{\mathrm{w}}n_{\mathrm{w}}(\mathbf{x_{\mathrm{w}}}-\mathbf{v}t,t)\psi_\alpha^\dagger(z,t) d^{\alpha\beta}\psi_\beta(z,t)\frac{\partial}{\partial z}V^{\mathrm{C}}(z_{\mathrm{ex}}-z_{\mathrm{w}},\rho_{\mathrm{w}},a)\\
    &=\int dz_{\mathrm{ex}} \int d{\mathbf{x}}_{\mathrm{w}}n_{\mathrm{w}}(\mathbf{x_{\mathrm{w}}}-\mathbf{v}t,t)n^{\alpha\beta}_{\mathrm{ex}}(z,t)d^{\alpha\beta}\frac{\partial}{\partial z}V^{\mathrm{C}}(z_{\mathrm{ex}}-z_{\mathrm{w}},\rho_{\mathrm{w}},a),
    \end{aligned}
\end{equation}
where $n^{\alpha\beta}=\psi^\dagger_{\alpha}\psi_\beta$ is the density of excitons.

In the following, we work at the RPA level and employ the Keldysh formalism~\cite{kavokineFluctuationinducedQuantumFriction2022}. We define the (bare) susceptibilies of water and excitons, as well as water-exciton correlator function as
\begin{equation}
    \chi_{\mathrm{w}}(\mathbf{x}_1,t_1, \mathrm{x}_2 t_2)=-\frac{i}{\hbar}\Big\langle \mathbb{T}_c \;n_{\mathrm{w}}(\mathbf{x}_1, t_1) n_{\mathrm{w}}(\mathbf{x}_2,t_2)\Big\rangle,
\end{equation}
\begin{equation}
    \chi_{\mathrm{ex}}^{\alpha\beta\gamma\delta}(z_1,t_1, z_2, t_2)=-\frac{i}{\hbar}\Big\langle \mathbb{T}_c \;n_{\mathrm{ex}}^{\alpha\beta}(z_1, t_1) n_{\mathrm{ex}}^{\gamma\delta}(z_2,t_2)\Big\rangle,
\end{equation}
\begin{equation}
    \chi_{\mathrm{ex-w}}^{\alpha\beta}(z,t_1, \mathbf{x}, t_2)=-\frac{i}{\hbar}\Big\langle \mathbb{T}_c \;n_{\mathrm{ex}}^{\alpha\beta}(z_1, t_1) n_{\mathrm{w}}(\mathbf{x},t_2)\Big\rangle,
\end{equation}
where $t_1,t_2\in c=c_1\cup c_2$ belong on the Keldysh contour that goes from negative to positive infinity ($c_1$) and back ($c_2$), and $\mathrm{T}_c$ is the time-ordering operator along this contour. As usual, we split the information in $\chi$ into a matrix of susceptibilities taking $t\in \mathbb{R}$ as argument,
\begin{equation}
    \chi_{\mathrm{w}}=\begin{pmatrix}
        \chi_\mathrm{w}^{11} & \chi_{\mathrm{w}}^{12} \\\chi_{\mathrm{w}}^{21}&\chi_{\mathrm{w}}^{22}
    \end{pmatrix}\quad \text{etc}.
\end{equation}
Of these, only three are independent, and we pass to the usual retarded, advanced, and Keldysh components:
\begin{equation}
    \begin{pmatrix}
        \chi^{\mathrm{R}} & \chi^{\mathrm{K}} \\0&\chi^{\mathrm{A}}
    \end{pmatrix}=\frac12\begingroup
\setlength\arraycolsep{10pt}\begin{pmatrix}
        \chi^{11}-\chi^{12}+\chi^{21}-\chi^{22}& \chi^{11}+\chi^{12}+\chi^{21}+\chi^{22}\\
        \chi^{11}-\chi^{12}-\chi^{21}+\chi^{22}& \chi^{11}+\chi^{12}-\chi^{21}-\chi^{22}
    \end{pmatrix}\endgroup.
\end{equation}
This matrix structure is respected by the perturbative expansion in the interaction Hamiltonian, i.e., water-exciton vertices are obtained by matrix multiplication and an appropriate convolution. Specifically, to the leading order in perturbation theory, to which we limit the analysis,
\begin{equation}\label{eq:supplementary_convolution}
\begin{aligned}
    \chi_{\mathrm{ex-w}}^{\alpha\beta}(z_{\mathrm{ex}},t_{\mathrm{ex}},\mathbf{x}_{\mathrm{w}}, t_{\mathrm{w}})=&\int_{-\infty}^{\infty} dt\int_{-\infty}^{\infty} dz_{\mathrm{ex}}'\int d\mathbf{x}_{\mathrm{w}}' \chi_{\mathrm{ex}}^{\alpha\beta\gamma\delta}(z_{\mathrm{ex}},t_{\mathrm{ex}},z_{\mathrm{ex}}',t)\chi_{\mathrm{w}}(\mathbf{x}_{\mathrm{w}}'-\mathbf{v}t,t,\mathbf{x}_{\mathrm{w}},t_{\mathrm{w}})\dots\\
    &d^{\gamma\delta}\frac{\partial}{\partial z}V^{\mathrm{C}}(z_{\mathrm{ex}}-z_{\mathrm{w}},\rho_{\mathrm{w}},a),
\end{aligned}
\end{equation}
where summation over repeated excitonic indices $\alpha,\beta$ and matrix multiplication over Keldysh indices is assumed. To simplify this equation, we introduce the water response function as in Eq.~\eqref{eq:supplementary_force_z_static}. The exciton confined response function should be defined in Fourier space as (suppressing the Keldysh indices)
\begin{equation}
    g_{\mathrm{ex}}(q,\omega)=-\frac{e^2}{4\pi\epsilon_0}(iqd^{\alpha\beta})(-iqd^{\gamma\delta})\cdot2I_0(qa)K_0(qa)\chi_{\mathrm{ex}}^{\alpha\beta\gamma\delta}(q,\omega),
\end{equation}
which ensures that it is dimensionless and has a positive imaginary component. The exciton-water confined response function should be defined as
\begin{equation}
    g_{\mathrm{ex-w}}(q,\omega)=-\frac{e^2}{4\pi\epsilon_0}(-iqd^{\alpha\beta}) \int dA'\frac{K_0(q\rho)}{K_0(qa)}\chi_{\mathrm{ex-w}}^{\alpha\beta}(q,\omega, \rho_{\mathrm{w}}).
\end{equation}
Under this definition, convolution~\eqref{eq:supplementary_convolution} becomes (Keldysh matrix) multiplication:
\begin{equation}
    g_{\mathrm{ex-w}}(q,\omega)=-g_{\mathrm{ex}}(q,\omega)g_{\mathrm{w}}(q,\omega-qv).
\end{equation}
Resolved in Keldysh components, this can be solved as
\begin{equation}\label{eq:supplementary_dyson_first_term}
    \begin{aligned}
        g_{\mathrm{ex-w}}^{\mathrm{R/A}}=-g_{\mathrm{ex}}^{\mathrm{R/A}}g_{\mathrm{w}}^{\mathrm{R/A}},\\
        g_{\mathrm{ex-w}}^{\mathrm{K}}=-g_{\mathrm{ex}}^{\mathrm{R}}g_{\mathrm{w}}^{\mathrm{K}}-g_{\mathrm{ex}}^{\mathrm{K}}g_{\mathrm{w}}^{\mathrm{A}}.
    \end{aligned}
\end{equation}

\subsection{Friction coefficient for a nanotube}

The latter of these equations we can use directly to evaluate the rate of momentum exchange between the excitons and water. In real space, the expression for the force is given by same-time correlator function, which is related to the Keldysh component of $\chi_{\mathrm{ex-w}}$:
\begin{equation}
    \begin{aligned}
    \langle F_{z}\rangle &=-\int dz_{\mathrm{ex}} \int d\mathbf{x}_{\mathrm{w}}\frac{\partial^2}{\partial z^2}V^{\mathrm{C}}(z_{\mathrm{ex}}-z_{\mathrm{w}},\rho_{\mathrm{w}},a) d^{\alpha\beta}\langle n_{\mathrm{ex}}^{\alpha\beta}(z_{\mathrm{ex}},t) n_{\mathrm{w}}(\mathbf{x}_{\mathrm{w}},t)\rangle\\
    &=-\int dz_{\mathrm{ex}} \int d\mathbf{x}_{\mathrm{w}}\frac{\partial^2}{\partial z^2}V^{\mathrm{C}}(z_{\mathrm{ex}}-z_{\mathrm{w}},\rho_{\mathrm{w}},a) d^{\alpha\beta}\;\frac{i\hbar}{2}\chi_{\mathrm{ex-w}}^{\mathrm{K},\alpha\beta}(z_{\mathrm{ex}},t,\mathbf{x}_{\mathrm{w}},t).
    \end{aligned}
\end{equation}
After a Fourier transform, we can derive
\begin{equation}
    \langle F_z\rangle=-\frac{\hbar L}{2}\int\frac{dq}{2\pi}\int \frac{d\omega}{2\pi}\; q\; g_{\mathrm{ex-w}}^{\mathrm{K}}(q,\omega),
\end{equation}
where $L=\int dz$ is the total length of the tube. 

After this point, the derivation follows the same steps as~\cite{kavokineFluctuationinducedQuantumFriction2022}. We substitute~\ref{eq:supplementary_dyson_first_term}, express the Keldysh components of response functions in terms of the retarded ones via the fluctuation-dissipation theorem. We then use parity properties of the response functions to pass to an integral involving the imaginary parts of the response functions only. Lastly, we expand it to linear order in velocity $v$ to obtain the friction coefficient:
\begin{equation}\label{eq:supplementary_lambda_quantum}
    \lambda_{\rm ex}^{\rm dyn}\!=\frac{\langle F_z\rangle}{2\pi aLv}=\!\frac{\hbar^2}{4\pi^3 k_\mathrm{B}T a}\!\int \!dq\; q^2 \!\!\int \! d\omega \frac{{\mathrm{Im}{g}_{\mathrm{ex}}^\mathrm{R}(q,\omega)\;\mathrm{Im}{g}_{\mathrm{w}}^\mathrm{R}}(q,\omega)}{\sinh^2(\hbar\omega/2k_\mathrm{B}T)}.
\end{equation}

\subsection{Exciton response function on a nanotube}
To compute the exciton response function in the case of a narrow nanotube characterized by a radius much smaller than $a_{\mathrm{B}}$, we first treat the excitons as a one-dimensional gas and evaluate its susceptibility defined as
\begin{equation}
    \chi_{\mathrm{ex}}^{\alpha\beta\gamma\delta}(z-z', t-t')=-\frac{i}{\hbar}\langle n_{\mathrm{ex}}^{\alpha \beta}(z,t) n_{\mathrm{ex}}^{\gamma\delta}(z',t')\rangle,
\end{equation}
and then compute the confined exciton response function (summation convention)
\begin{align}
    g_{\mathrm{ex}}(q,\omega)&=-\frac{e^2}{4\pi\epsilon_0}(iqd^{\alpha\beta})(-iqd^{\gamma\delta})\cdot2I_0(qa)K_0(qa)\chi_{\mathrm{ex}}^{\alpha\beta\gamma\delta}(q,\omega)\\
    &=-\frac{e^2}{4\pi\epsilon_0}q^2\;2I_0(qa)K_0(qa)\;d^{\alpha\beta}d^{\gamma\delta}\chi_{\mathrm{ex}}^{\alpha\beta\gamma\delta}(q,\omega)\\
    &\approx -\frac{e^2}{4\pi\epsilon_0}q^2\;\frac{1}{qa}\;d^{\alpha\beta}d^{\gamma\delta}\chi_{\mathrm{ex}}^{\alpha\beta\gamma\delta}(q,\omega),
\end{align}
where in the last step we have used the planar approximation.

The calculation of susceptibility $\chi_{\mathrm{ex}}$ is similar to that found in~\cite{Kamenev2011}. For non-interacting excitons, the susceptibility is given by a bubble diagram:
\begin{equation}\label{eq:supplementary_susceptibility_in_terms_of_G}
    \chi_\mathrm{ex}^{\mathrm{R}}(1,2)=\frac{i\hbar}{2}[G^{\mathrm{K}}(1,2)G^{\mathrm{A}}(2,1)+G^{\mathrm{R}}(1,2)G^{\mathrm{K}}(2,1)],
\end{equation}
where we have restored the Keldysh indices $A$, $R$, $K$ and introduced free exciton Green's functions. Since the excitons are non-interacting, they cannot transition between internal states under free motion and we may write (without summation convention)
\begin{equation}
    G^{\mathrm{A,R,K}}_{\alpha\beta}=\delta_{\alpha\beta}G^{\mathrm{A,R,K}}_{\alpha}
\end{equation}

Explicit expressions for these components in the Fourier space can be found within the assumption of quasi-equilibrium~\cite{Kamenev2011}:
\begin{align}
    G^{\mathrm{R}}_\alpha (q,\omega)&=\frac{1}{\hbar\omega-\mathcal{E}^{\alpha}_q+ i0}\\
    G^{\mathrm{A}}_\alpha (q,\omega)&=\frac{1}{\hbar\omega-\mathcal{E}^{\alpha}_q- i0}\\
    \begin{split}
    G^{\mathrm{K}}_\alpha (q,\omega)&=2if(\hbar\omega-\mu)\mathrm{Im}(G^{\mathrm{R}}_\alpha (q,\omega))\\
    &=-{2\pi i}\;\delta(\hbar\omega-\mathcal{E}^{\alpha}_q)f(\hbar\omega-\mu)
    \end{split},
\end{align}
where
\begin{equation}
    f(\hbar\omega-\mu)=\coth\left(\frac{\hbar\omega-\mu}{2k_{\mathrm{B}}T}\right)=2n_{\mathrm{B}}(\hbar\omega)-1,
\end{equation}
where $\mu$ is the chemical potential of excitons that controls their linear density, and $n_{\mathrm{B}}$ is the Bose--Einstein distribution function. We can substitute these results into Eq.~\eqref{eq:supplementary_susceptibility_in_terms_of_G} transforming to the Fourier space:
\begin{equation}\label{eq:supplementary_susceptibility}
\begin{aligned}
    \left(\chi_{\mathrm{ex}}^{\mathrm{R}}\right)^{\alpha\beta\gamma\delta}&=\frac{i\hbar}{2}\delta_{\beta\gamma}\delta_{\alpha\delta}\int\frac{dq'}{2\pi}\frac{d\omega'}{2\pi}[G^{\mathrm{R}}_\beta(q+q',\omega+\omega')G^{\mathrm{K}}_\alpha(q',\omega')\\
    &\qquad +G^{\mathrm{K}}_\beta(q+q',\omega+\omega')G^{\mathrm{A}}_\alpha(q',\omega')]\\
    &=\frac{i\hbar}{2}(-2\pi i)\delta_{\beta\gamma}\delta_{\alpha\delta}\int\frac{dq'}{2\pi}\frac{d\omega'}{2\pi}\Biggr[\frac{f(\hbar\omega')\delta(\hbar\omega'-\mathcal{E}^{\alpha}_{q'})}{\hbar\omega+\hbar\omega'-\mathcal{E}^{\beta}_{q+q'}+i0}\\
    &\qquad+\frac{f(\hbar\omega+\hbar\omega')\delta(\hbar\omega+\hbar\omega'-\mathcal{E}^{\beta}_{q+q'})}{\hbar\omega'-\mathcal{E}^{\alpha}_{q'}-i0}\Biggr]\\
    &=\frac{1}{2}\delta_{\beta\gamma}\delta_{\alpha\delta}\int\frac{dq'}{2\pi}\Biggr[\frac{f(\mathcal{E}^{\alpha}_{q'})}{\hbar\omega+\mathcal{E}_{\alpha}(q')-\mathcal{E}^{\beta}_{q+q'}+i0}\\
    &\qquad+\frac{f(\mathcal{E}^{\beta}_{q+q'})}{-\hbar\omega+\mathcal{E}^{\beta}_{q+q'}-\mathcal{E}^{\alpha}_{q'}-i0}\Biggr]\\&=\delta_{\beta\gamma}\delta_{\alpha\delta}\int\frac{dq'}{2\pi}\frac{n_{\mathrm{B}}(\mathcal{E}^{\alpha}_{q'})-n_{\mathrm{B}}(\mathcal{E}^{\beta}_{q+q'})}{\hbar\omega+\mathcal{E}^{\alpha}_{q'}-\mathcal{E}^{\beta}_{q+q'}+i0}\\ &=\delta_{\beta\gamma}\delta_{\alpha\delta}\int\frac{dq'}{2\pi}\frac{n_{\mathrm{B}}(\mathcal{E}^{\alpha}_{q'})-n_{\mathrm{B}}(\mathcal{E}^{\beta}_{q+q'})}{\hbar\omega+E_{\alpha}-E_{\beta}-\frac{\hbar^2q^2}{2M}-\frac{\hbar^2 q}{M}q'+i0}
\end{aligned}
\end{equation}

We now want to express $n_{\mathrm{B}}$ in terms of the linear density of excitons $n_{\mathrm{ex}}^{\rm 1D}$. The normalization condition is
\begin{equation}\label{eq:supplementary_normalization}
    \sum_\alpha\int\frac{dq}{2\pi} n_{\mathrm{B}}(\mathcal{E}^\alpha_{q})=n_{\mathrm{ex}}^{\rm 1D}.
\end{equation}

To solve this equation analytically, we make a simplifying assumption $n_{\mathrm{B}}\ll1$ which amounts to replacing the Bose-Einstein distribution with the Boltzmann distribution:
\begin{equation}
    n_{\mathrm{B}}(E)=\frac{1}{e^{(E-\mu)/k_{\mathrm{B}}T}-1}\approx \mathrm{const}\cdot e^{-E/k_{\mathrm{B}}T}
\end{equation}
Substitution of Eq.~\eqref{eq:supplementary_exciton_dispersion_supplementary} into Eq.~\eqref{eq:supplementary_normalization} yields
\begin{equation}
    n_{\mathrm{B}}(\mathcal{E}^\alpha_{q})=\frac{n_{\mathrm{1D}}}{\mathcal{Z}}\exp\left(-\frac{\mathcal{E}^\alpha_q}{k_{\mathrm{B}}T}\right)=\frac{2\sqrt{\pi}}{q_{\mathrm{th}}}n_{\mathrm{ex}}^{\rm 1D}\frac{\exp({-E_{\alpha}/k_{\mathrm{B}}T})}{\sum_\beta \exp({-E_{\beta}/k_{\mathrm{B}}T})}\exp\left(-\frac{q^2}{q_{\mathrm{th}}^2}\right),
\end{equation}
where $\mathcal{Z}=\sum_\alpha\int\frac{dq}{2\pi}\exp(-\mathcal{E}^\alpha_q/k_{\mathrm{B}}T)$ is the partition function, $q_{\mathrm{th}}=\sqrt{2Mk_{\mathrm{B}}T/\hbar^2}$ is the thermal wavenumber. This assumption $n_{\mathrm{B}}\ll1$ applies for $n_{\mathrm{ex}}^{\rm 1D}\ll q_{\mathrm{th}}$. For example, at room temperature if $M=m_{\mathrm{e}}$, this yields $n_{\mathrm{ex}}^{\rm 1D}\ll\qty{1}{\per\nano\meter}$.

We now restrict ourselves to a specific model of excitons accounting for only two internal eigenstates with an allowed dipole transition. We set the energy of the lower (ground) state to zero, and the excited state has energy $\Delta E$. We denote the corresponding transitional dipole moment as $d^{01}\ne 0$. Since we are only interested in the imaginary part of $g_{\mathrm{ex}}$, we may use the Sokhotski--Plemelj theorem in the form
\begin{equation}
    \mathrm{Im} \int \frac{dq'}{2\pi}\frac{F(q')}{q_0-q'-i0}=-\frac12 F(q_0),
\end{equation}
where $F$ is a test function. We can thus find the following closed form expression:
\begin{equation}\label{eq:supplementary_g_ex}
\begin{aligned}
\mathrm{Im}\,g_{\mathrm{ex}}(q,\omega)&=\frac{e^2}{4\pi\epsilon_0}q^2\;2I_0(qa)K_0(qa)\;|d^{01}|^2\cdot \frac{2\sqrt{\pi}}{q_{\mathrm{th}}}\frac{n_{\mathrm{ex}}^{\rm 1D}}{1+\exp(-\Delta E/k_{\mathrm{B}}T)}\frac{M}{\hbar^2 q}\frac12\\
    &\qquad \times\left[e^{-q_-^2/q_{\mathrm{th}}^2}-e^{-\Delta E/k_{\mathrm{B}}T}e^{-(q_- +q)^2/q_{\mathrm{th}}^2}+e^{-\Delta E/k_{\mathrm{B}}T}e^{-q_+^2/q_{\mathrm{th}}^2} - e^{-(q_+ +q)^2/q_{\mathrm{th}}^2}\right]\\
    &=\frac{e^2}{4\pi\epsilon_0}q\;2I_0(qa)K_0(qa)\;|d^{01}|^2 \frac{n_{\mathrm{ex}}^{\rm 1D}}{1+\exp(-\Delta E/k_{\mathrm{B}}T)} \sqrt{\frac{\pi M}{2 \hbar^2 k_{\mathrm{B}}T}}\\
    \\
    &\qquad \times\left[e^{-q_-^2/q_{\mathrm{th}}^2}-e^{-\Delta E/k_{\mathrm{B}}T}e^{-(q_- +q)^2/q_{\mathrm{th}}^2}+e^{-\Delta E/k_{\mathrm{B}}T}e^{-q_+^2/q_{\mathrm{th}}^2} - e^{-(q_+ +q)^2/q_{\mathrm{th}}^2}\right]
\end{aligned}
\end{equation}
where
\begin{equation}
    q_{\pm}=-\frac{q}{2}+\frac{M}{\hbar^2 q}(\hbar \omega \pm \Delta E)
\end{equation}
are the poles of the integral in Eq.~\eqref{eq:supplementary_susceptibility}. Within the planar approximation, the factor $2 I_0(qa) K_0(qa)$ should be consistently replaced with $1/qa$.

The dominant term in the final bracket of Eq.~\eqref{eq:supplementary_g_ex} is usually the first one, while the rest are typically smaller than one. The first term peaks when $q_-=0$ corresponding to $\hbar\omega=\Delta E+\hbar^2 q^2/2M=\mathcal{E}_q$. The width of the response function around this dispersion relation is set by the temperature of the system.

Conceptually, the response function Eq.~\eqref{eq:supplementary_g_ex} accounts for two types of fluctuations that populate the excited states of the bound electron-hole system. First, these are the thermal fluctuations, which are responsible for the factors of $\exp(-\Delta E/k_{\mathrm{B}}T)$ in the equation. In addition to the thermal fluctuations, the quantum fluctuations are accounted for. Indeed, the transition energy is also implicitly encoded in $q_{\pm}$ which appears in the equation. And even in the limit $\Delta E\to \infty$, when thermal population of the excited states is exponentially low, $g_{\mathrm{ex}}$ exhibits peaks in the vicinity of $(q,\omega)$ that satisfy $q_-=0$. This means that excited states remain populated even in the low temperature limit, as excitons virtually transition to them as a result of the electrostatic interaction with other particles or excitations such as hydrons in our case. Hence, this equation accounts for both thermal and quantum fluctuations of excitons.

In terms of scalings with excitonic mass $M$, we see that $\mathrm{Im}\,g_{\mathrm{ex}}^{\mathrm{R}}\sim \sqrt{M}$. When integrated for the friction coefficient in Eq.~\eqref{eq:supplementary_lambda_quantum}, the relevant region involves values of $q$ for which $\hbar q^2/2 M\lesssim \omega_{\mathrm{D}}$, the Debye frequency of water. Hence, in the expression for the friction coefficient, $q^2 dq\sim M^{3/2}$ and overall $\lambda_{\mathrm{ex}}^{\mathrm{dyn}}\sim M^2$. In terms of the scalings with energy gap $\Delta E$, if it is large, the exciton response function is exponentially suppressed in the region of water response frequencies. And so the friction is also exponentially suppressed at high $\Delta E$.

\subsection{Static excitons as a limiting case of dynamic excitons}
It can be shown that the model of static excitons can be formally obtained from the general results for dynamic excitons as a limiting case. For that, we need to assume a large exciton mass $M$ and set $\Delta E=0$. This is equivalent to considering only the ground state but allowing it to have a non-zero static dipole moment $\braket{0|\hat{d}_z|0}=d$, which is the case for excitons trapped in external potential. In this case, Eq.~\eqref{eq:supplementary_g_ex} becomes

\begin{equation}
    g_{\mathrm{ex}}(q,\omega)=\frac{e^2}{4\pi\epsilon_0}q\;2I_0(qa)K_0(qa)\;d^2 n_{\mathrm{ex}}^{\rm 1D}\sqrt{\frac{\pi M}{2 \hbar^2 k_{\mathrm{B}}T}}\left[e^{-q_-^2/q_{\mathrm{th}}^2}-e^{-(q_- +q)^2/q_{\mathrm{th}}^2}\right]
\end{equation}

Examine the last factor using the definition of $q_{-}$:
\begin{equation}
    e^{-q_-^2/q_{\mathrm{th}}^2}-e^{-(q_- +q)^2/q_{\mathrm{th}}^2}=2\exp\left[-\left(\frac{M\omega}{\hbar q q_{\mathrm{th}}}\right)^2\right]\exp\left[-\left(\frac{q}{2 q_{\mathrm{th}}}\right)^2\right]\sinh \frac{M \omega}{\hbar q_{\mathrm{th}^2}}
\end{equation}
Recall that in the fluctuation-induced friction Eq.~\eqref{eq:supplementary_lambda_quantum}, we are integrating against the product $\mathrm{Im}\,g_{\mathrm{ex}}^{\mathrm{R}}\;\mathrm{Im}\,g_{\mathrm{w}}^{\mathrm{R}}$. The latter factor decays exponentially at large $q$, so we are considering only $q\lesssim\qty{20}{\per\nano\meter}$~\cite{kavokineFluctuationinducedQuantumFriction2022}. So, as we consider excitons with higher and higher mass, the second exponential factor becomes closer to 1 for all considered wavevectors. Meanwhile, the first exponential factor decays for $\omega \gtrsim \hbar q q_{\mathrm{th}}/M$. Notice that the argument of the hyperbolic sine is then small, as $q_{\mathrm{th}}\gg q$ for sufficiently large mass $M$. Thus, we can approximate it linearly. Overall, we get the following expression:

\begin{equation}
    g_{\mathrm{ex}}(q,\omega)=\frac{e^2}{4\pi\epsilon_0}q\;2I_0(qa)K_0(qa)\;d^2 n_{\mathrm{ex}}^{\rm 1D}\sqrt{\frac{\pi M}{2 \hbar^2 k_{\mathrm{B}}T}}\times \frac{\hbar \omega}{k_{\mathrm{B}}T}\;\exp\left[-\left(\frac{M\omega}{\hbar q q_{\mathrm{th}}}\right)^2\right].
\end{equation}

Now we can substitute this into the expression for the dynamic excitonic friction Eq.~\eqref{eq:supplementary_lambda_quantum}. As mentioned before, the excitonic wavefunction is most pronounced at $\omega\lesssim \hbar q q_{\mathrm{th}}/M$, and this range becomes more narrow with an increase of $M$. Hence, we are interested in the local behaviour of the integrand around $\omega=0$. Recall that $\mathrm{Im}\,g_{\mathrm{w}}^{\mathrm{R}}(q,0)=0$, so we expand the response function of water to linear order in frequency. Similarly, we expand the hyperbolic sine to leading order:

\begin{equation}
\begin{aligned}
    \lambda_{\rm ex}^{\rm dyn}&=\frac{\hbar^2}{4\pi^3 k_\mathrm{B}T a}\frac{e^2}{4\pi \epsilon_0}d^2 n_{\mathrm{ex}}^{\rm 1D}\sqrt{\frac{\pi M}{2 \hbar^2 k_{\mathrm{B}}T}}\!\int_0^\infty \!dq\; q^3 \;2I_0(qa)K_0(qa)\dots\\
    &\qquad \dots \int_0^\infty \! d\omega \frac{\hbar\omega/k_{\mathrm{B}}T}{(\hbar\omega/2k_\mathrm{B}T)^2}
    \exp\left[-\left(\frac{M\omega}{\hbar q q_{\mathrm{th}}}\right)^2\right]\;\omega\frac{\partial}{\partial \omega}\mathrm{Im}\,{g}_{\mathrm{w}}^\mathrm{R}(q,0)\\
    &=\frac{\hbar^2}{4\pi^3 k_\mathrm{B}T a}\frac{e^2}{4\pi \epsilon_0}d^2 n_{\mathrm{ex}}^{\rm 1D}\sqrt{\frac{\pi M}{2 \hbar^2 k_{\mathrm{B}}T}}\sqrt{\frac{\pi k_{\mathrm{B}}T}{2 M }}\frac{4k_{\mathrm{B}}T}{\hbar}\\
    &\qquad \times\int_0^\infty dq\,q^4 2 I_0(qa)K_0(qa)\partial_\omega \mathrm{Im}\,{g}_{\mathrm{w}}^\mathrm{R}(q,0)\\
    &=\frac{e^2}{4\pi\epsilon_0}\frac{n_{\mathrm{ex}}^{\rm 1D}}{\pi a}\int_0^\infty dq\; (qd)^2 q^2 I_0(qa)K_0(qa)\partial_\omega \mathrm{Im}\,{g}_{\mathrm{w}}^\mathrm{R}(q,0)\\
    &=\lambda_{\mathrm{ex}}^{\mathrm{stat}},
\end{aligned}
\end{equation}
where we have identified $|\Lambda(q)|=qd$ for the point-like electric dipole model of dynamic excitons that we have assumed.

\begin{figure}[ht]
    \centering
    \captionsetup{justification=centering}
    \includegraphics[width=0.5\linewidth]{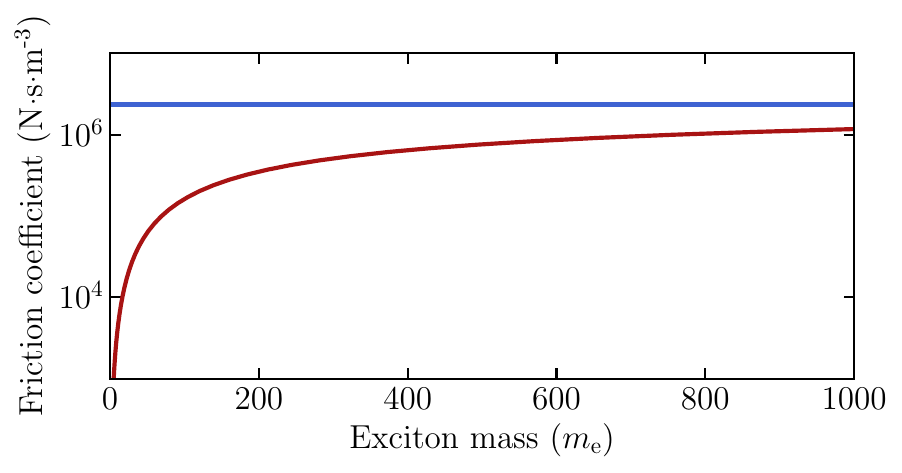}
    \caption{\response{Friction coefficient of dynamic excitons with polarized ground state plotted against the excitonic mass $M$ (red). Dipole moment $d=\qty{1}{\nano\meter}$, concentration $n_{\mathrm{ex}}^{\mathrm{1D}}=\qty{0.1}{\per\nano\meter}$, nanotube radius $a=\qty{0.39}{\nano\meter}$. Friction coefficient of static excitons with the same dipole moment and concentration (Blue).}}
    \label{fig:SM_lambda_vs_M}
\end{figure}

To summarize, the dynamic model of excitons does include static excitons as a limiting case when $M\to\infty$ and $\Delta E\to 0$. A specific condition for excitons to be approximately static is $q_{\mathrm{th}}\gtrsim q_{\mathrm{w}}$, where $q_{\mathrm{w}}\sim \qty{30}{\per\nano\meter}$ is the characteristic wavevector of water response at $\omega=0$. At room temperature this translates to $M\gtrsim 10^3\,m_{\mathrm{e}}$. \response{Indeed, the numerical calculations on Fig.~\ref{fig:SM_lambda_vs_M} demonstrate that as the excitonic mass increases to large values, the friction converges to that of static excitons.}

This calculation can be regarded as a consistency check of the developed framework. It also justifies thinking of static excitons as free excitons with a large mass. This will be used in comparison of our model with the experimental results on diffusion of nanotubes.

\subsection{Exciton response function and excitonic friction in 2D}

In this section, we briefly extend our results to excitons in 2D semiconducting materials instead of 1D nanotubes. The equation for the friction coefficient in this case is the same as in~\cite{kavokineFluctuationinducedQuantumFriction2022}, because that framework applies to all Coulomb-interacting media with a planar interface.:
\begin{equation}
    \lambda_{\mathrm{ex}}^{\mathrm{2D}}=\frac{\hbar^2}{8\pi^2 k_{\mathrm{B}}T}\int_0^\infty dq\;q^3 \int_0^\infty d\omega\;\frac{1}{\sinh^2(\hbar\omega/2k_{\mathrm{B}}T)}\mathrm{Im}\,g_{\mathrm{ex}}^{\mathrm{R}}(q,\omega)\;\mathrm{Im}\,g_{\mathrm{w}}^{\mathrm{R}}(q,\omega).
\end{equation}
The expression for the response function of excitons must be specified:
\begin{equation}
    g_{\text{ex, 2D}}^{\mathrm{R}}=-\frac{e^2}{4\pi\epsilon_0}\frac{2\pi}{q}(\mathbf{d}^{\alpha\beta}\cdot\mathbf{q})(\mathbf{d}^{\gamma\delta}\cdot\mathbf{q})\left(\chi_{\mathrm{ex, 2D}}^{\mathrm{R}}\right)^{\alpha\beta\gamma\delta},
\end{equation}
where
\begin{equation}
    \left(\chi_{\mathrm{ex, 2D}}^{\mathrm{R}}\right)^{\alpha\beta\gamma\delta}(\mathbf{q},\omega)=\delta_{\beta\gamma}\delta_{\alpha\delta}\int\frac{d\mathbf{q}'}{(2\pi)^2}\frac{n_{\mathrm{B}}(\mathcal{E}^{\alpha}_{\mathbf{q}'})-n_{\mathrm{B}}(\mathcal{E}^{\beta}_{\mathbf{q}+\mathbf{q}'})}{\hbar\omega+E_{\alpha}-E_{\beta}-\frac{\hbar^2\mathbf{q}^2}{2M}-\frac{\hbar^2}{M}\mathbf{q}\cdot \mathbf{q}'+i0},
\end{equation}
and the integration is taken over the 2D plane $(q_x,q_y)$. If we again assume Boltzmann distribution, we can perform the integral analytically and obtain the expression with the same notation as before:
\begin{equation}
\begin{aligned}
    \mathrm{Im}\,g_{\mathrm{ex, 2D}}^{\mathrm{R}}(q,\omega)&=\frac{e^2}{4\pi\epsilon_0}\;|d^{01}|^2 \frac{n_{\mathrm{ex}}^{2D}}{1+\exp(-\Delta E/k_{\mathrm{B}}T)} \sqrt{\frac{2\pi^3 M}{ \hbar^2 k_{\mathrm{B}}T}}\\
    \\
    &\times\left[e^{-q_-^2/q_{\mathrm{th}}^2}-e^{-\Delta E/k_{\mathrm{B}}T}e^{-(q_- +q)^2/q_{\mathrm{th}}^2}+e^{-\Delta E/k_{\mathrm{B}}T}e^{-q_+^2/q_{\mathrm{th}}^2} - e^{-(q_+ +q)^2/q_{\mathrm{th}}^2}\right].
\end{aligned}
\end{equation}

Physically speaking, 2D excitons should correspond to the limit of nanotubes with large radius. However, the expressions derived in this and the previous section cannot be formally obtained from each other. This is because we have assumed that the nanotubes are sufficiently narrow so that the excitonic density is radially symmetric. Taking into an account modes with nonzero angular momentum and the corresponding Bessel functions would indeed allow to show that the limit of large radius matches with the case of a 2D interface.

The general conclusion about peak of the response function around $\hbar\omega=\mathcal{E}_q$ and scaling of the friction with $\Delta E$ are the same as for one-dimensional excitons. The scaling with exciton mass is slightly altered, as the friction coefficient now has $q^3 dq\sim M^2$, so $\lambda_{\mathrm{ex,2D}}^{\rm dyn}\sim M^{5/2}$ instead of $M^2$.

\subsection{Emergence of static dipole moment}
In this subsection, we provide a simple two-state model that explains how external electrostatic potential due to, for example, organic corona around a nanotube can induce a static dipole moment in the ground state. Consider a pure nanotube with two excitonic eigenstates, which we call $\ket{1s}$ and $\ket{2s}$. They represent the one-dimensional relative motion wavefunction of an exciton. Due to reflective symmetry of the nanotube, they do not have a static dipole moment, but a non-zero transitional dipole moment is possible and scales like $\braket{1s|\hat{d}|2s}=a\sim a_\mathrm{B}$. Now suppose that the electronic system of the nanotube is perturbed by the presence of an external potential, e.g., due to charge imbalances in the organic corona. This creates off-diagonal terms in the Hamiltonian, which we call $U$. Overall, we may write the Hamiltonian and dipole moment in matrix notation as
\begin{equation}
    \hat{\mathcal{H}}=\begin{pmatrix}
        0 & U\\
        U & \Delta E
    \end{pmatrix}\quad \hat{d}=\begin{pmatrix}
        0& a\\ a&0
    \end{pmatrix}.
\end{equation}
This simple model can be solved analytically for the perturbed ground state $\ket{1s'}$ and the static dipole moment in this state:
\begin{equation}
    \ket{1s'}=\begin{pmatrix}
        \cos\frac{\theta}{2}\\-\sin\frac{\theta}{2}
    \end{pmatrix}
\end{equation}
\begin{equation}
    d=\braket{1s'|\hat{d}|1s'}=-a\sin\theta=-\frac{2Ua}{\sqrt{\Delta E^2+4U^2}},
\end{equation}
where $\theta=\arccos(1/\sqrt{1+4U^2/\Delta E^2})$. For small $U$, $d\approx -2Ua/\Delta E$. If we assume $a\sim \qty{2}{\nano\meter}$, $\Delta E\sim \qty{200}{\milli\eV}$, and $U\sim \qty{25}{\milli\eV}$, we get $d\sim \qty{0.5}{\nano\meter}$, which is already at the saturation value of the static dipole moment model. Thus, even $U\ll \Delta E$ is sufficient to create a strong dipole moment.

\section{Hydrodynamic drag coefficient and application to the experiment of Kistwal et al.}
In this section we derive the formula for the diffusion of nanotubes with finite slip length~$b$ within slender body theory (SBT) and discuss the effects of finite Reynolds numbers. Finally, we compare our results with the measurements of Kistwal et al.
\subsection{Drag formulas from slender body theory without slip}
Let a cylinder of length $L$ and radius $a$ be submerged in a viscous liquids. There are two classical expressions that describe the motion of high aspect ratio cylinders in a slow viscous flow. One of them assumes an infinitely long cylinder ($L/a\to \infty$) and finite but small Reynolds number
\begin{equation}
    \Re =\frac{a U}{\nu}\ll1,
\end{equation}
where $\nu$ is the kinematic viscosity of the fluid, and $U$ is the flow velocity. This formula reads~\cite{Landau1987Fluid}:
\begin{equation}
    F_\perp=\frac{4\pi \eta LU_\perp}{\log\frac{1}{\Re}+\frac12-\gamma_{\mathrm{E}}-2\ln2+O(\log^{-1}\frac1\Re)},
\end{equation}
where $\eta$ is the dynamics viscosity of the fluid, $\gamma_{\mathrm{E}}\approx 0.577$ is the Euler--Mascheroni constant, and the subscript $\perp$ indicates flow perpendicular to the axis of the cylinder. The expression for the resistance to translations along the axis of the cylinder cannot be derived in this limit, because it diverges with $L/a\to \infty$

A different expression is derived from the slender body theory in slow viscous flow. It assumes an identically zero Reynold's number ($\Re\to0$), but finite aspect ratio
\begin{equation}
    \frac{L}{a}\gg 1.
\end{equation}

The results of the slender body theory can be summarized by writing the components of the drag matrix $\xi$ for translation parallel  ($\parallel$) and normal ($\perp$) to the axis of the cylinder:
\begin{align}
    \label{eq:supplementary_SBP_perp}\xi_\perp^{-1}=\frac{U_{\perp}}{F_{\perp}} = \frac{1}{4\pi \eta L}\left(\log\frac{L}{a}+\gamma_{\perp}\right),\\
    \label{eq:supplementary_SBT_parallel}\xi_\parallel^{-1}=\frac{U_{\parallel}}{F_{\parallel}} = \frac{1}{2\pi \eta L}\left(\log\frac{L}{a}+\gamma_{\parallel}\right),
\end{align}
where $\gamma_\perp$ and $\gamma_\parallel$ are $O(1)$ numbers that depend weakly on the aspect ratio $L/a$ and approach certain values (less than one) for infinite cylinder. For sufficiently long carbon nanotubes, therefore, these corrections are dominated by the log-terms and can be neglected. 

\subsection{Drag formulas from slender body theory including slippage}
The above expressions were derived under the assumption of no-slip boundary condition. Nevertheless, finite slip length is known to have an important effect on the fluid dynamics of CNTs. Below we derive the results for the drag coefficients $\xi_\parallel$, $\xi_\perp$ assuming creeping flow equations $\Re=0$ and finite slip length. We begin with the drag for parallel translations. As the cylinder translates along its axis, the flow locally around it is flowing in the same ($z$) direction. The Navier--Stokes equation in cylindrical coordinates for velocity $\mathbf{u}=(0,0,w(r))$ then reads simply
\begin{equation}\nabla^2 w=0.\end{equation}
The finite slip boundary condition is 
\begin{equation}U_\parallel-w(a)=b \frac{\mathrm{d}w}{\mathrm{d}r}(a),\end{equation}
where $U_{\parallel}$ is the translational velocity of the cylinder. The general solution is
\begin{equation}
    w(r)=U_\parallel \frac{\log \frac{r^*}{r}}{\log\frac{r^*}{a}+\frac{b}{a}},
\end{equation}
where $r^*$ is an arbitrary constant, such that $w(r^*)=0$. As usual for the 2D Stokes equations, this solution contains a logarithmic divergence, which must be taken care of. The physically relevant cut-off is $r\sim L$, which is a scale at which the fluid no longer flows only along the tube. Therefore, we may with \emph{logarithmic accuracy} set $r^*=L/2$ in the above solution. The choice of the factor of $1/2$ is conventional and usually results in small sub-logarithmic corrections~\cite{tiradoTranslationalFrictionCoefficients1979}. From here, the net force on the tube is easily computed from the viscous stress $\eta w'(a)$ and we obtain

\begin{equation}\label{eq:supplementary_SBT_parallel_slip}
    \xi^{-1}_{\parallel}=\frac{U_\parallel}{F_\parallel}=\frac{1}{2\pi \eta L}\left(\log\frac{L}{2a}+\frac{b}{a}+O(1)\right),
\end{equation}
which generalizes Eq.~\eqref{eq:supplementary_SBT_parallel} to the case of finite slip. From the above equation we see that the slip length becomes relevant when $b\gtrsim a$. In the limit $b\to \infty$ (perfect slip), this equation formally predicts zero drag force, because this analysis neglects the flow structure near cylinder ends. In that region the length scale of the flow near ends is $a$, so the force contribution from that region is
\begin{equation}F_\parallel^{\text{ends}} \sim \eta a U_\parallel,\end{equation}
regardless of the value of slip length. This becomes comparable with the prediction of Eq.~\eqref{eq:supplementary_SBT_parallel_slip} for $b\sim L$. Thus, we will assume $b\ll L$ for our calculations.

We proceed to the calculation of drag for the translation normal to the axis of the cylinder. For the moment, we will assume that the slip length $b$ is the same as for the case of tangential motion. The general solution to the 2D Stokes equation that diverges no faster than logarithmically at infinity and depends linearly on the velocity of the cylinder $\mathbf{U}_\perp$ comes from Papkovich--Neuberg potentials~\cite{TRANCONG198272}
\begin{equation}\label{eq:supplementary_papkovich}\chi=\alpha a^2\mathbf{U}_\perp\cdot\nabla \log\frac{r}{a},\quad \bm{\Phi} = \beta \mathbf{U}_\perp \log\frac{r}{a}+\gamma\mathbf{U}_\perp,\end{equation}
where $\alpha$ and $\beta$ are constants to be determined and $r^*$ is again a cutoff radius. For these potentials the flow field is
\begin{equation}\mathbf{u}=\mathbf{U}_\perp\left(-\beta  \log\frac{r}{a}+\alpha \frac{a^2}{r^2}-\gamma\right)+\frac{1}{r^2}(\mathbf{U}_\perp\cdot\mathbf{x})\mathbf{x}\left(\beta-2\alpha\frac{a^2}{r^2}\right).\end{equation}

In this parametrisation, the vector potential $\bm{\Phi}$ represents the 2D Stokeslet component of the solution, which is the only component contributing to the net force per unit length of the cylinder as
\begin{equation}\frac{\mathbf{F}_\perp}{L} = {4\pi \eta \beta}\mathbf{U}_\perp\end{equation}

The boundary conditions at $r=a$ are 
\begin{equation}\mathbf{u}\cdot \mathbf{x}=\mathbf{U}_\perp \cdot\mathbf{x}\end{equation}
\begin{equation}(1-\mathbf{n}\mathbf{n})\cdot(\mathbf{u}-\mathbf{U})=b (1-\mathbf{n}\mathbf{n})(\nabla\mathbf{u}+(\nabla\mathbf{u})^T)\cdot\mathbf{n},\end{equation}
where $\mathbf{n}=\mathbf{x}/r$ is the unit normal vector to the cylinder. We must also carefully specify the condition at infinity. For large radii, the velocity field becomes
\begin{equation}\mathbf{u}=\mathbf{U}_\perp\left(\beta\log \frac{r}{a}-\gamma\right)+O\left(\frac1{r^2}\right).\end{equation}
To get an answer with a logarithmic density, we can impose the same regularization boundary condition as for the case of parallel translations:
\begin{equation}\label{eq:supplementary_regularisation}\mathbf{u}\left(r=\frac{L}{2}\right)=0\end{equation}

Overall, we get a system of equations on the unknown constants in Eq.~\eqref{eq:supplementary_papkovich}:
\begin{equation}\begin{cases}
    \beta-\alpha-\gamma = 1\\
    \alpha-\gamma-1=-4\alpha\frac{b}{a}\\
    \gamma=\beta\log\frac{L}{2a},
\end{cases}\end{equation}
From which we get
\begin{equation}\beta=\frac{-1}{\log\frac{L}{2a}-\frac{a+4b}{2a+4b}}\end{equation}
This appears similar to Eq.~\eqref{eq:supplementary_SBT_parallel_slip}, but there is an important difference: the new term in the denominator doesn't grow with $b$, but stays bounded between 1/2 and 1. Since this analysis neglects the constant terms compared to the logarithmic ones (logarithmic accuracy), here the extra term in the denominator must be dropped for consistency. Observe that this implies that we do not have to assume that the slip length for the flow normal to the axis of the nanotube experiences the same slip length as the flow parallel to the axis. Overall, we get again
\begin{equation}\label{eq:supplementary_SBT_perp_slip}
    \xi^{-1}_{\perp}=\frac{U_\perp}{F_\perp}=\frac{1}{4\pi \eta L}\left(\log\frac{L}{2a}+O(1)\right),
\end{equation}
In other words, slip length only modifies the tangential drag coefficient significantly. We can write our results in a matrix notation
\begin{equation}\label{eq:supplementary_SBT_slip_matrix}
    \mathbf{F}=-\frac{2\pi\eta L}{\log\frac{L}{2a}}\begin{pmatrix}
    \frac{1}{1+b/a\log(L/2a)}& 0 & 0\\
    0& 2 & 0\\
    0 & 0 & 2
    \end{pmatrix} 	\mathbf{U}
\end{equation}

\subsection{Diffusion coefficient}
If a particle has a linear drag force $\mathbf{F}=-\xi\mathbf{U}$, where $\xi$ is a matrix, and the orientation of the particle is random, then the average diffusion coefficient is given by~\cite{Landau1987Fluid}
\begin{equation}
    D=\frac{1}{3}k_{\text{B}}T\Tr( \xi^{-1}).
\end{equation}

Applying this to Eq.~\eqref{eq:supplementary_SBT_slip_matrix}, we obtain
\begin{equation}\label{eq:supplementary_diffusion}
    D=\frac{k_{\mathrm{B}}T}{3\pi\eta L}\left[\log\left(\frac{L}{2a}\right)+\gamma+\frac{b}{2a}\right],
\end{equation}
where we have absorbed all $O(1)$ corrections discussed above into the quantity $\gamma$. The last term is a novel result that encompasses the effect of finite slip length on the diffusion. The larger the slip length, the smaller the friction and drag, and thus, the nanotube diffuses more easily.

In the limit $b/a=0$, the value of $\gamma$ is taken as $\gamma\approx 0.32$~\cite{tiradoTranslationalFrictionCoefficients1979}; however, this value depends slightly on the shape of cylinder's ends~\cite{aragonHighPrecisionTransport2009}. Accounting for slip length, $\gamma$ also depends on the ratio $b/a$. Moreover, it depends on the slip length for flow normal to the nanotube, which does not necessary match with the slip length for parallel flow. Nevertheless, we expect $|\gamma|\lesssim 1$ in all cases.

If $b\gg a\log(L/2a)$, the formula simplifies significantly and formally matches with the equation for diffusion of spherical particles with radius $La/b$:
\begin{equation}D=\frac{k_{\text{B}}Tb}{6\pi\eta La}
\end{equation}

\subsection{Finite Reynolds number effects}

The derivation of the previous sections assumed that the inertial terms of the Navier--Stokes equation are negligible, and the cutoff scale is set by the cylinder length $L$. However, if the velocity of the rod is sufficiently high, there is another cutoff scale $a/\mathrm{Re}$. When both of these length are treated as large but finite, one must use the smaller one for regularization. In the case of parallel drag $\xi_\parallel$, however, the regularization scale is always $r^*\sim L$, because the inertial terms vanish identically for the unidirectional axisymmetric flow~\cite{Landau1987Fluid}. For the perpendicular component of the drag there is thus a competition between these two scales which we estimate for typical carbon nanotubes. The relevant velocity scale is set by the thermal motion of CNTs:
\begin{equation}v_{\text{th}}\sim \sqrt{\frac{k_{\text{B}}T}M},\end{equation}
where $M$ is the mass of the CNT given by $M=2\pi a L \rho_{\text{2D}}$, where $\rho_{\text{2D}}=\qty{6.6e-7}{\kilo\gram\per\meter\squared}$ is the surface density of graphene. We then compute the ratio
$$\frac{L/a}{1/\mathrm{Re}}\sim\frac{L}{a}\frac{v_\text{th}a}{\nu}\sim \frac{1}{\nu}\left(\frac{k_{\text{B}}T}{\rho_{\text{2D}}}\right)^{1/2}\times \left(\frac{L}{a}\right)^{1/2}\sim 0.1\left(\frac{L}{a}\right)^{1/2}$$

This result is rather interesting, because it shows that for aspect ratios $10\lesssim L/a\lesssim 1000$ both cutoff scales are comparable. For very long CNT with $L/a\gg 1000$, the relevant cutoff scale is $a/\mathrm{Re}$, while for CNT of modest lengths $L/a\ll 10$ it is $L$ (although the slender body theory still requires $L/a\gg 1$). For the case of experiment in~\cite{Kistwal2026}, the aspect ratio is $L/a\sim 10^3$, so $\log(L/2a)\approx \log(1/\Re)$ with logarithmic accuracy and both are smaller than the slip-length term $b/a$ in Eq.~\eqref{eq:supplementary_diffusion}




\subsection{Comparing with the experiment of Kistwal et al.}
To compare the developed theory of excitonic friction to the experimental results of Kistwal et al.~\cite{Kistwal2026}, we must determine how the incident power is related to the excitonic density. The nanotubes used for the main experiment were (6,5)-enriched SWCNTs, functionalized with DNA molecules. For diffusion measurements, the fluorescence correlation spectroscopy (FCS) technique is used with a $\qty{485}{\nano\meter}$ laser with photon energy $E_{\mathrm{ph}}=\qty{2.6}{\eV}$. In (6,5) SWCNT's this corresponds approximately to absorption peak related to the $E_{22}$ excitonic formation, so we assume that absorbed photons generate $E_{22}$ exciton-hole pairs which bind into excitons. The corresponding orientation-averaged light absorption cross section is $\sigma_{\mathrm{C}}=0.6-\qty{1.0e-17}{\centi\meter\squared}$ per carbon atom~\cite{berciaudLuminescenceDecayAbsorption2008, streitDirectlyMeasuredOptical2014}. We use the value $\sigma_{\mathrm{C}}=\qty{e-17}{\centi\meter\squared}$. The absorption cross-section per unit length of the nanotube is obtained by multiplying $\sigma_{\mathrm{C}}$ with number of carbon atoms per unit length $\rho_{\mathrm{C}}$. From geometry, for (6,5) nanotubes, it is $\rho_{\mathrm{C}}=\qty{88.4}{\per\nano\meter}$~\cite{sanchezSpecificAbsorptionCross2016}.

The power in the experiment is not focused on a single nanotube, but instead it is localized in a confocal volume $V_{\mathrm{cf}}=\pi^{3/2}w_{xy}^2 w_z =\qty{1.5}{\femto\liter}$. Using the reported values of diffusion constant and diffusion time we can deduce from the article $w_{xy}\approx\qty{300}{\nano\meter}$ and hence $w_z\approx\qty{3} {\micro\meter}$. We can define the confocal cross section as $A_{\mathrm{cf}}=\pi w_{xy}^2=\qty{0.28}{\micro\meter\squared}$. Therefore, the fraction $N\sigma_{\mathrm{C}}/A_{\mathrm{cf}}$ of total power $P$ is absorbed by unit length of a nanotube in the confocal volume. The rate of generation of electron-hole pairs is thus $P\rho_{\mathrm{C}}\sigma_{\mathrm{C}}/A_{\mathrm{cf}}E_{\mathrm{ph}}$. For example, if $P=\qty{10}{\micro\watt}$ and a nanotube of length $L=\qty{600}{\nano\meter}$, the rate of photon absorption is $\qty{4.5e9}{\per \second}$. 

Very quickly after excitation, the electron-hole pair drops into the excitonic state of lowest energy and then stays in this state for a long time until recombining. The equilibrium linear density of excitons $n_{\mathrm{ex}}^{\rm 1D}$ is set by the balance of creation and annihilation. To quantify the rate of annihilation, the mean photoluminescence (PL) lifetime of excitons $\tau_{\mathrm{ex}}$ is usually introduced, and we discuss its value below. The dynamical equation for $n_{\mathrm{ex}}^{\rm 1D}$ is
\begin{equation}
    \frac{d{n}_{\mathrm{ex}}^{\mathrm{1D}}}{dt}=\frac{P \rho_{\mathrm{C}} \sigma_{\mathrm{C}}}{A_{\mathrm{cf}}E_{\mathrm{ph}}}-\frac{n_{\mathrm{ex}}^{\rm 1D}}{\tau_{\mathrm{ex}}}.
\end{equation}
In the steady state, we can obtain the power-to-density ratio of excitons:
\begin{equation}\label{eq:supplementary_power_to_density}
    \frac{P}{n_{\mathrm{ex}}^{\rm 1D}}=\frac{A_{\mathrm{cf}}E_{\mathrm{ph}}}{\rho_{\mathrm{C}} \sigma_{\mathrm{C}}\tau_{\mathrm{ex}}}=\frac{1300}{\tau_{\mathrm{ex}}[\unit{ns}]}\frac{\unit{\micro\watt}}{\unit{\per\nano\meter}}
\end{equation}

The lifetime of excitons $\tau_{\mathrm{ex}}$ is determined by radiative and non-radiative processes usually involving phonons. Radiative lifetime is usually very large, reaching infinite values for excitons outside the light cone~\cite{spataruTheoryInitioCalculation2005}. In most situations, including the paper considered, the quantum yield, which is the measure of emitted photons per absorbed photon, is low, corresponding to mostly non-radiative decay of excitons~\cite{miyauchiRadiativeLifetimesCoherence2009}. Estimates of lifetime of excitons vary a lot in the literature. In pure nanotubes it is on the order of $10-\qty{100}{\pico\second}$~\cite{perebeinosRadiativeLifetimeExcitons2005, spataruTheoryInitioCalculation2005, birkmeierProbingUltrafastDynamics2022}. However, it was determined that chemical functionalization and localizing defects can significantly increase the lifetime of excitons into the nanosecond range~\cite{hartmannPhotoluminescenceDynamicsAryl2016a, zhengPhotoluminescenceDynamicsDefined2021, changQuantumDefectsCarbon2026} reaching up to $\qty{1.5}{\nano\second}$~\cite{heIntrinsicLimitsDefectstate2019}.

Another model parameter to discuss is the non-excitonic slip length. We must distinguish the friction outside and inside the nanotube, which differ by orders of magnitude for small-radius nanotubes, as described in~\cite{falkMolecularOriginFast2010a}. Here, the effect on diffusion is controlled by the outside friction, which is larger for narrower nanotubes. For the nanotubes considered here, the friction coefficient is $\lambda_0\approx\qty{2e4}{\newton\second\per\meter\cubed}$, corresponding to $b_0\approx\qty{50}{\nano\meter}$~\cite{falkMolecularOriginFast2010a}. We will use this value while keeping in mind that the DNA functionalization may decrease the slip length by introducing surface roughness.

We now discuss the appropriate choice of model for the description of the experiment. Specifically, we argue that the correct choice is of static excitons. The reason for this is twofold. First, in all experiments nanotubes are wrapped with organic molecules for stability, which creates an external potential for the excitons. One of the effects of this potential is to create a binding effect to certain preferable sites thus significantly increasing the effective mass of the exciton as a whole. The second effect, as outlined in section 2.6 of this SI is the emergence of a static dipole moment due to mixing of eigenstates. Together, these effects are sufficient to make the approximation of treating excitons as static, as described above in section 2.4 of the SI.

We also offer an explanation for the observed reduction in the effect of excitonic friction when the $sp^3$ defects are added to the nanotube lattice. These defects localize the excitons symmetrically, without a static dipole moment~\cite{zhengPhotoluminescenceDynamicsDefined2021}. Therefore, these excitons should be treated as quantum dots with friction scaling like $\exp(-\Delta E/k_{\mathrm{B}}T)$, where $\Delta E$ is the transition energy to the first excited state. Thus, even though these excitons have an infinite mass, they do not possess static dipole moment, and thus exhibit weak friction. Moreover, measurement of photoluminescence of nanotubes with defects shows that these localized excitons emit light more strongly~\cite{Kistwal2026}, which was previously attributed to a reversed bright-dark state ordering~\cite{hartmannPhotoluminescenceDynamicsAryl2016a} of these excitons. Therefore, at the same applied power $P$ the average density of excitons lower for nanotubes with $sp^3$ defects, which further decreases the effect of excitonic friciton.

\subsection{Estimating factors that can reduce diffusion}
\response{

The absolute value of diffusion, calculated using Eq.~\eqref{eq:supplementary_diffusion}, overestimate the values measured experimentally, regardless of the way excitons are modeled. This can be demonstrated by considering two extreme cases. First, the classical no-slip condition, which has been assumed in previous studies of CNT diffusion, obtained by setting $b=0$. This corresponds to ``infinite friction'', and correspondingly, the lowest possible value of diffusion within the hydrodynamic framework. The computed value is $D_{\text{no slip}}=\qty{4.9}{\micro\meter\squared\per\second}$. This must be compared with the lowest friction coefficient measured in any experiment of~\citet{Kistwal2026}, which is $\qty{0.9}{\micro\meter\squared\per\second}$. Similar results were previously obtained in experiments on CNT diffusion~~\cite{tsyboulskiTranslationalRotationalDynamics2008, streitMeasuringSingleWalledCarbon2012}. In these works the authors usually attribute the discrepancy to the ``wall-effect'' due to the specific geometry of experiment, where nanotubes diffuse in a constrained space. The standard approach to account for this observed difference is to introduce a fitting phenomenological constant factor in Eq.~\eqref{eq:supplementary_diffusion}. A wall-effect cannot be ruled out from the experiment of~\citet{Kistwal2026}; however, it is also possible that several other factors can complement the explanation of the difference between the absolute value of predicted and measured diffusion. 

Having performed the new analytic calculations of diffusion for the case of finite slip length, which has been repeatedly observed in nanotubes, we may extend this analysis to consider the opposite limiting case, when no excitons are present in the nanotube, and the slip length takes its usual value of $\qty{50}{\nano\meter}$~\cite{falkMolecularOriginFast2010a}. In this case Eq.~\eqref{eq:supplementary_diffusion} predicts $D_{\text{no exciton}}=\qty{50}{\micro\meter\squared\per\second}$, much larger than the value extracted from fitting the experiment to the non-excitonic regime. Both of the considered limits demonstrate that the overestimation of diffusion by Eq.~\eqref{eq:supplementary_diffusion} is not due to the treatment of excitons, but instead should be explained by other factors, which we consider and estimate below.

The first factor to consider is the length of the nanotubes. The reported length of $\qty{600}{\nano\meter}$ clearly represents the sample average or median length, and nanotubes are known to exhibit a wide distribution of length with a long tail at large lengths~\cite{streitMeasuringSingleWalledCarbon2012}. The technique of measuring diffusion (FCS) relies on detection of re-emitted light by the nanotubes. Longer nanotubes absorb and emit more light, and therefore, the measurements are skewed towards the diffusion values of longer nanotubes, which are correspondingly smaller. As a rough estimate, we may assume that the nanotube lengths follow a normal distribution with mean length $L$ and standard deviation $\sigma\sim L$. The nanotube's rate of absorption and emission is proportional to its length, so the distribution must be weighted by a linear factor in length, giving the probability density function $f(x)\sim x\exp(-(x-L)^2/2\sigma^2)$. Using this distribution, the mean length actually measured in experiment would be
$L+\sigma^2/L$, which is significantly larger than $L$. In this case the measurement is significantly biased by the longest nanotubes in the sample, and the diffusion measurements are correspondingly smaller. This can be corrected by introducing an effective length $L^*$ larger than $L$. If we disregard the logarithmic dependence in Eq.~\eqref{eq:supplementary_diffusion}, this is mathematically equivalent to introducing a phenomenological prefactor.

Meanwhile, the length of the nanotube does not influence the predicted excitonic friction and slip length. Indeed, the concentration of excitons on the nanotubes is independent  of the length of the latter, as can be seen from Eq.~\eqref{eq:supplementary_power_to_density}. Therefore, all nanotubes in the sample have roughly the same value of the slip length despite differences in length.

The next factor to consider is the radius of the nanotube. For a pure SWCNT the diameter is a purely geometric quantity computed from the hexagonal lattice of graphene. For a chirality index $(m,n)$ the diameter is~\cite{LIU200575}
\begin{equation}\frac{a_0}{\pi}\sqrt{m^2+mn+n^2},\end{equation}
where $a_0=\qty{0.246}{\nano\meter}$ is the lattice constant. Nevertheless, this geometric radius is not necessarily the right quantity to use for hydrodynamic calculations. This is partially because of a transition layer near the nanotube, where the water molecules are repelled by Lennard-Jones forces, leading to almost zero density within the first $\qty{0.2}{\nano\meter}$~\cite{waltherHydrodynamicPropertiesCarbon2004}. Additionally, the effective radius is increased by the organic functionalization around the nanotube, increasing the radius by $\qty{0.5}{\nano\meter}$ according to estimates~\cite{campbellAtomicForceMicroscopy2008, kodeInteractionDNAComplexedBoron2021}. Thus, this factor can potentially increase the effective radius by a factor of 2 or more. In Eq.~\eqref{eq:supplementary_diffusion} this affects mostly the term involving slip length, which is inversely proportional to the radius. However, for large slip observed in CNT's this term generally dominates, and thus, this effect can again be accounted by an overal phenomenological prefactor.

Concerning the effect of increased hydrodynamic radius on the excitonic friction, we must distinguished the two sources mentioned above. The van der Waals repulsion is already accounted in the response function of water, obtained from numerical simulations, in which water molecules interact with carbon atoms. Therefore, this correction must not be considered when calculating the excitonic friction, and the geometric radius should be used, corresponding to the fact that electrons and holes are distributed in close vicinity from the carbon shell of the nanotube. The presence of organic functionalization can affect the excitonic friction by changing the electric properties of both excitons and water. The effect of functionalization on excitons has been discussed above in Sec. 2.6, where it was demonstrated that even small potential gradients due to organic charges can polarize the ground state of the exciton. Moreover, such gradients can act as a disorder potential, increasing the effective mass or even localizing the excitons~\cite{singh2017ExcitonLocalization}. Both of these effects make the friction stronger, bringing it into the regime of static excitons. 

The third possible contribution to the discrepancy may be the effect of high nanotube concentration.
Estimating its impact on the effective viscosity is similar to the Einstein's calculation of viscosity of suspensions of spherical particles~\cite{Landau1987Fluid}. However, the elongated shape of carbon nanotubes modifies the traditional picture. The standard approach is to employ the Kirkwood--Auer--Batchelor (KAB) formula~\cite{kodeInteractionDNAComplexedBoron2021, davisPhaseBehaviorRheology2004, batchelorSlenderbodyTheoryParticles1970, kirkwoodViscoElasticPropertiesSolutions1951a}, which gives to the leading order for very slender bodies
\begin{equation}\label{eq:supplementary_effective_viscosity}
    \eta=\eta_0\left(1+\frac{2}{45}\frac{L^2/a^2}{\ln(L/a)}\phi\right)=\eta_0\left(1+\frac{2\pi}{45}\frac{L^3}{\ln(L/a)}n_{\mathrm{3D}}\right),
\end{equation}
where $\eta_0$ is the bulk viscosity, $\phi=\pi a^2 L n_{\mathrm{3D}}$ is the volume fraction of nanotubes in the solution, and $n_{\mathrm{3D}}$ is the concentration of nanotubes. In this equation, we have neglected the sub-leading logarithmic corrections for the purpose of estimating the effect. This formula is also only valid in the linear regime, when the correction term is smaller than one. In the experiment the concentration used is $1\;\mathrm{nM}$, corresponding to $n_{\mathrm{3D}}=\qty{0.6}{\per\cubic\micro\meter}$. Assuming the non-corrected values of $L=\qty{600}{\nano\meter}$, $a=\qty{0.39}{\nano\meter}$, we obtain the correction of 0.25\%, which is rather small. It is important to note, however, that larger lengths of nanotubes can increase it due to cubic dependence on $L$. Moreover, the employed formula was derived under no-slip assumptions, and it is possible that slip may significantly change the prediction. 

Beyond the impact on effective viscosity, the high concentration may also modify the diffusion coefficient of individual CNTs due to steric or hydrodynamic flow interactions. In particular, the overlap between flows generated by multiple CNTs is challenging to capture, but probably impacts the diffusion coefficient more than suggested by the estimated viscosity. Therefore, we cannot rule out potential contribution of steric effects on the absolute value of diffusion.

Importantly, the effect of high concentration of nanotubes cannot affect the excitonic friction. In the calculation of the latter, the shape of the large-scale hydrodynamic profile is irrelevant, and only the local flow velocity enters the equations. The excitonic state is also unaffected by the presence of nanotubes, because the mean distance between nanotubes at such concentrations is $~\qty{1}{\micro\meter}$, which is much larger than the spacial extent of the exciton's wavefunction and the typical lengthscale of electrostatic interaction with water, both of which are typically only a few nanometers. Overall, hydrodynamic interactions between CNT's appear irrelevant to the prediction of nanotube diffusion for the specific experiment, although it may become important for setups with longer nanotubes and higher concentrations.

}

\bibliographystyle{plainnat}
\bibliography{bibliography, bibfile}